\documentclass[12pt]{article}

\usepackage{graphics}
\usepackage{epsfig}
\usepackage{epsfig}
\newcommand{\Py}{{\sc Pythia}}

\newcommand{\Ar}{{\sc Ariadne}}

\newcommand{\Hw}{{\sc Herwig}}
\newcommand{\q}{\mathrm{q}}
\newcommand{\qbar}{\mathrm{\overline{q}}}
\newcommand{\g}{\mathrm{g}}
\newcommand{\LQCD}{\Lambda_{QCD}}
\newlength{\captivewidth}
\newcommand{\captive}[1]{\rule{5mm}{0mm}%
\begin{minipage}{\captivewidth}%
\caption[small]{#1}\end{minipage}}
\def\Journal#1#2#3#4{{#1}{\bf #2} (#3) #4}

\def\NPB{{\it Nucl. Phys.~}{\bf B}}
\def\PLB{{\it Phys. Lett.~}{\bf  B}}

\def\ZPC{{\it Z. Phys.~}{\bf C}}

\def\JHEP{\it J. High Energy Phys.~}

\def\CPC{\it Computer Phys. Commun.~}
\def\PRP{\it Phys. Rept.~}

\def\EPJC{{\it Eur. Phys. J.~}{\bf C}}

\begin{document}
\begin{titlepage}
\begin{flushright}
  LU TP 02-13 \\
  July 2002
\end{flushright}
\vspace{2cm}
\begin{center}
  \Large
  {\bf The $\lambda$-Measure and the Generalised Dipoles in the
        Lund Model} \\
  \normalsize \vspace{2mm}
  Bo Andersson \\
  Sandipan Mohanty\footnote{sandipan@thep.lu.se}  \\
  Fredrik S\"oderberg\footnote{fredrik@thep.lu.se} \\

  Department of Theoretical Physics, Lund University, \\
  S\"olvegatan 14A, S-223 62 Lund, Sweden
\end{center}
\vspace{1cm}
\noindent {\bf Abstract} 
We demonstrate  that the  multiple gluon emission  phase space  in the
dipole cascade model  has a strong linear correlation  with the number
of gluons emitted. The number of gluons per unit available phase space
at a certain resolution scale is found to be remarkably independent of
the  cms energy  and global  event  properties like  thrust, and  even
changes in the ordering variable or resolution scale. We show that the
distribution of sizes  of gluon-gluon dipoles in a  parton cascade has
stability properties which are sufficient  to account for those of the
phase space  variable. We observe that certain  more abstract entities,
 defined  in the  context of  hadronisation  and related  to the  gluon
emission phase space, share those properties of colour dipoles and name
them  Generalised Dipoles.   We also  present an  analytical  model to
qualitatively describe our findings.
\end{titlepage}


\setcounter{page}{1}
\section{Introduction}
The phase space for multiple bremsstrahlung emission of photons in QED
is  entirely given  by the  properties of  the original  current.  The
reason  is that  the  photon-quanta are  uncharged  and therefore  the
currents are  not changed because  of emission of  photons (besides the
recoil effects from hard emissions).

On  the other  hand the  QCD field  quanta, the  gluons,  are charged.
Already  the emission of  a first  QCD gluon  in $e^+e^-$-annihilation
means that  the original  ${\q} {\qbar}$ dipole  is changed.   It so
happens that  the change  is (to a  very good approximation)  from one
dipole to two independent dipoles,  one between the quark $({\q})$ and
the first  gluon and  the other between  the gluon and  the anti-quark
$({\qbar})$.  The initial  dipole is at rest in  the total cms whereas
the two ``new'' dipoles move  away from each other. The combined phase
space for  emitting from either one  or the other of  these dipoles is
found to be  larger.  In rapidity space the  increase can be described
as an  extra region of  a size corresponding  to the logarithm  of the
squared transverse momentum of the first emitted gluon.

In the Dipole  Cascade Model \cite{GGUP}, as implemented  in the event
generator  {\Ar}   \cite{LL},  a   second  emission  leads   to  three
independent  dipoles   and  so  on.   The  ordering   variable  is  an
invariantly defined  transverse momentum  of the emitted  gluons.  The
dipole masses decrease quickly with successive emissions.

{\Hw} \cite{PMBWII}, and  {\Py} \cite{TS},  subdivide the
emerging  dipole  angular  regions  into independent  cones  for  each
emitting parton.   One way to  state the coherence conditions  is that
they do not permit double-counting in the emission process.  {\Hw} and
{\Py} implement the coherence conditions  by means of angular
ordering to avoid  such overlaps.  The ordering variable  for {\Hw} is
just  the angle  (or the rapidity  variable) occurring  in the
coherence conditions. {\Py} uses the ``virtuality'' along the emission
lines  as  an ordering  variable  and  only  afterwards introduces  an
angular ordering condition.

In this paper  we will mainly describe the  emerging features in terms
of the notions of the Dipole  Cascade Model ({\Ar}) and only perform a
cursory  comparison to  similar results  from  {\Py}.  Our  aim is  to
investigate  a set  of  distributions stemming  from the  perturbative
parton cascades.   We will  show that the  cascades result in  a local
structure, corresponding to sets of independent entities, which we are
going  to call  Generalised  Dipoles ($GD$s).   The  $GD$s are  linked
together along the  colour lines of the QCD force  field.  They have a
common distribution in a  generalised rapidity range $\lambda$, and in
the (local) transverse momentum.  This structure is independent of the
total cms  energy, the global event variables  like thrust 
and  the  number  of  hard   emissions.   Further  the  $GD$s  show  a
surprisingly   small   scale  dependence,   i.e.    almost  the   same
distributions occur inside a wide range of the ordering variable.

We will concentrate upon $e^+e^-$-annihilation events, where the gluon
cascade is produced through  the brems\-strah\-lung from an originally
produced  ${\q}{\qbar}$  pair  (although   we  expect  that  the  same
structures will  emerge also in  other dynamical situations).   Such a
multi-gluon state is conveniently  described by means of a four-vector
valued function, the  directrix $\mathcal{A}_{\mu}$ \cite{Artru}.  The
directrix  is obtained  by  laying out  the energy--momentum  vectors,
$k_j$,  of the  emitted  partons in  colour-order,  starting with  the
${\q}$ and ending  with the ${\qbar}$ (we will  neglect the production
of ``new'' ${\q} {\qbar}$-pairs  through the gluon splitting process).
As  we will  use massless  partons, the  directrix is  a curve  with a
tangent that is lightlike everywhere.

As mentioned earlier the phase space available for gluon emission at a
certain  value of  the ordering  variable $k_{\perp}$  depends  on the
partons  present and  their  colour order.   It  could, therefore,  be
regarded  as a  functional of  the directrix  $\mathcal{A}_{\mu}$.  In
section  \ref{dicascade}, we  will  discuss a  measure  of this  phase
space,  to   be  called  $\ell$,   for  multiple  gluon   emission  in
perturbative  QCD. An  infrared  stable generalisation  of this  phase
space, involving a ``resolution parameter'' $m_0$ leads to the concept
of the $\lambda$-measure \cite{BAGGBS}.

The $\lambda$-measure was introduced  many years ago \cite{BAGGBS}, as
a generalisation  of the rapidity  range, to describe the  phase space
for hadronisation. At the same  time \cite{BAGGBS}, a curve similar in
appearance to  a set of  connected hyperbolae along the  directrix was
introduced,  the  $\mathcal{  X}$-curve.   Its length  is  related  to
$\lambda$  in  the same  way  as  the  ``ordinary'' rapidity  variable
measures  the length  of  a hyperbola  spanned  between two  lightlike
vectors.

The  tangent  vector,   to  be  called  $q_T$,  at   a  point  on  the
$\mathcal{X}$-curve  that reaches  out  to a  point  on the  directrix
quickly  reaches a  constant  length, $q_T^2  =m_0^2$,  just like  the
constant distance  in Minkowski metric  between the lightcones  and an
ordinary hyperbola. The parameter  $m_0$ is a resolution parameter for
the properties of  the directrix. In the context  of hadronisation, it
is  fixed by  the  properties  of the  spectrum  of hadrons  produced.
Thought of as a generalisation  of $\ell$, the $\lambda$-measure has a
resolution  parameter which can  be directly  related to  the ordering
variable in  the perturbative cascade.  When the  tangent vector $q_T$
follows the directrix it will span a surface between the directrix and
the $\mathcal{X}$-curve.   The $\lambda$-measure is  also proportional
to the area of this surface.

In the  dipole model,  the available phase  space for  gluon emission,
$\ell$, in the Leading Log  Approximation (LLA) \cite{DKMT} and in the
Modified Leading Log Approximation (MLLA) \cite{DKMT} schemes is a sum
of  a series  of  terms ,  $(\Delta  \ell)_j$, one  for  each pair  of
consecutive  vectors along the  directrix \cite{BAGGBS}.   These terms
are contributions  to the phase  space from individual dipoles  in the
Dipole Cascade Model.  We will  show that the $\lambda$-measure can be
similarly subdivided into a  number of parts $(\Delta \lambda)_j$, one
for  each vector  along  the  directrix, and  that  they are  strongly
related to the contribution  from the dipoles, $(\Delta \ell)_j$.  The
process of subdividing $\lambda$  corresponds to a partitioning of the
surface mentioned above into ''plaquettes''\footnote{In this paper, we
will use the  word plaquette to denote this area  even though one part
of  the boundary  curve  is a  part  of a  hyperbola.  The  plaquettes
defined  in this  paper  are not  related  to the  plaquettes used  in
Lattice  QCD.}  that  are enclosed  by one  ``initial'' $q_T$  and one
``final''  $q_T$ vector  together  with a  hyperbolic  segment of  the
$\mathcal{X}$-curve and the  parton energy--momentum vector in between,
cf.  Fig \ref{deltalambdaarea} in  section \ref{dicascade}.  It is the
regularity in  the $(\Delta \lambda)$  distribution that will  lead to
the definition  of the  $GD$s.  For the  surface mentioned  above this
regularity  implies a  simple general  structure, which  shows  only a
slowly varying dependence on the ordering variable.

The directrix also plays a major  role in the description of the state
of the massless relativistic string, which  is used as a model for the
QCD force  fields in  String Fragmentation \cite{Artru}.   The surface
spanned by  the string during one  period is a  minimal surface.  This
has  two consequences.   On the  one  hand the  surface is  completely
determined by  its boundary  curve. On the  other hand the  surface is
stable against  small deformations (the model is  infrared stable). In
the motion of  a massless relativistic string, a  wave moves across the
string surface bouncing at the  endpoints defined by the orbits of the
${\q}$ and  the ${\qbar}$.  This  means that the  internal excitations
(which in the Lund Model are  identified as the gluons) will reach and
affect  the endpoints  in turn,  i.e.  in  the colour  order  of their
emission.  The  corresponding orbit  of  the  ${\q}$  endpoint is  the
directrix.

It is interesting  to note that the $\mathcal{  X}$-curve (or rather a
close relative, called the  $\mathcal{ P}$-curve in \cite{BASMFS}) and
the $\lambda$-measure  play a major  role in the  String Fragmentation
process for multigluon  string states. In the process  that we devised
in \cite{BASMFS},  the final state hadron energy--momenta,  laid out in
rank order, constitute a  curve, the $X$-curve.  The $X$-curve follows
the directrix at a typical stochastically fluctuating distance and the
area in between the two curves could  be thought of as the area in the
Lund  Model Area  law \cite{BASMFS}.   It turns  out that  the average
$X$-curve  is just the  $\mathcal{ P}$  curve with  a length  given by
$\lambda$.  The  generalised dipoles to  be defined in this  note also
play an  interesting role in  the fragmentation process. Just  as they
emerge as  independent entities from the partonic  cascades, they also
fragment essentially independently of  each other into the final state
hadrons, in a fragmentation scheme along the directrix.

The  $\lambda$-measure has  been used  in many  different  contexts in
investigations  over  the  years.   It  was quickly  found,  that  the
$\mathcal{X}$-curve   shows   properties   of  a   fractal   character
\cite{BAGGPD}  and  the  multi-fractal  dimensions were  shown  to  be
identical to the so-called anomalous dimensions of QCD \cite{GGAN}.

In section \ref{dicascade}, we recapitulate some features of the gluon
emission phase space $\ell$-measure and the $\lambda$-measure with the
aim  of pointing  out  similarities and  differences  between the  two
quantities\footnote{In the past, like in the references listed in this
paper,  the gluon  emission  phase  space has  often  been called  the
$\lambda$-measure. But since both this phase space and the phase space
for hadronisation are relevant for  this work, we make the distinction
here and give them different names.}. We also introduce the quantities
$\Delta \lambda$ and present  a geometrical interpretation for them.
In  section \ref{results},  we describe  our findings  and  define the
Generalised Dipoles.  In section  \ref{discussion}, we present a model
to understand the distributions we  have obtained. At the same time we
will  consider  some earlier  results  and  the theoretical  analysis.
Finally, in section \ref{conclu} we make some concluding remarks.


\section{The $\lambda$-Measure and The Perturbative Cascades}
\label{dicascade}
\subsection{The phase space in a dipole cascade and the $\ell$-Measure}

The probability for the emission of  gluons in the Dipole Model can be
described   in    terms   of   an   inclusive    density   of   gluons
(cf. \cite{GGUP}):
\begin{eqnarray}
\label{1gluon}
\mathrm{d}n =  \alpha_{\mathrm{eff}} dy \frac{dk_{\perp}^2}{k_{\perp}^2} \times
{\textstyle(\sum_p)}
\end{eqnarray} 
with  $\alpha_{\mathrm{eff}}$  the   effective  running  coupling  and
$(\sum_p)$ the  polarisation sum, i.e.  the coupling  between the spin
of the emitters and the  gluon.  The coherence conditions in this case
are identical to energy--momentum  conservation, i.e. the cms energy of
the gluon cannot exceed half the total cms energy
\begin{eqnarray}
\label{energymom}
k_{\perp} \cosh(y) \leq \frac{ \sqrt{s}}{2}
\end{eqnarray}
A   convenient  approximation,  corresponding   to  the   Leading  Log
Approximation,  is  $|y|<  \frac{1}{2} \ln(\frac{s}{k_{\perp}^2})$  so
that   the   total  rapidity   range   is   $\Delta   y  \simeq   \ln(
\frac{s}{k_{\perp}^2})$.   As we  mentioned  above the  emission of  a
gluon will  change the  current. But it  is an  immense simplification
that this change is simply a change from one dipole to two independent
dipoles  (to  a  very good  approximation)  \cite{DKMT,Azimov}.
Thus the  density for the emission  of two gluons  is factorisable, so
that we obtain
\begin{eqnarray}
\label{2gluon}
\mathrm{d}n({\q} \mathrm{ g_1g_2} {\qbar}) \simeq \mathrm{d}n({\q}
\mathrm{g_1}{\qbar})\left[\mathrm{d}n({\q}\mathrm{g_2g_1}) + 
\mathrm{d}n(\mathrm{g_1 g_2}{\qbar})\right]
\end{eqnarray}
(with a negative correction that formally is of the order of $1/N_c^2$
for a finite number of colours,  but reduced further in practice due to
kinematics).  Eq.   (\ref{2gluon}) is valid if  the transverse momenta
are ordered so that $k_{\perp 1} > k_{\perp 2}$ or else the two gluons
are exchanged. While the original $({\q}{\qbar})$ dipole is at rest in
the cms the two ``new'' dipoles move apart. Each of them will decay in
its own  restframe with a  phase space given by  Eq. (\ref{energymom})
and  this means  that the  combined rapidity  region for  emitting the
second gluon with transverse momentum $k_{\perp}$ is
\begin{eqnarray}
\label{2gluonphase}
(\Delta     \ell)_{01}+(\Delta    \ell)_{12}     &=&     \ln    \left(
\frac{s_{01}}{k_{\perp}^2}\right)         +         \ln         \left(
\frac{s_{12}}{k_{\perp}^2}\right)     \\     \nonumber     &=&     \ln
\left(\frac{s}{k_{\perp}^2}\right)+\ln\left(\frac{s_{01}s_{12}}{s
k_{\perp}^2}\right)
\end{eqnarray}
The expression $\frac{s_{01}s_{12}}{s}$  is the conventionally defined
invariant squared  transverse momentum  of the first  gluon, $k_{\perp
1}^2$.  Similarly, for  an emission at $k_{\perp}^2$ from  a system of
$n$ gluons, we can write the phase space as:
\begin{eqnarray}
\label{ngluonphase}
\ell  &=&  \sum_{i=0}^{n}(\Delta  \ell)_{i,i+1} =  \sum_{i=0}^{n}  \ln
(\frac{s_{i,i+1}}{k_{\perp}^2})       =       \ln      \prod_{i=0}^{n}
(\frac{s_{i,i+1}}{k_{\perp}^2})
\end{eqnarray}

At this  point, it is necessary  to differ between  two definitions of
the  transverse momentum.   The one  we  have used  up to  now is  the
ordering variable $k_{\perp}$, i.e. the value used for the emission of
a gluon and defined by the partitioning of a dipole of mass $M$ into a
pair    of     dipoles    with    masses     $(m_1,m_2)$    so    that
$k_{\perp}^2=m_1^2m_2^2/M^2$. There  is another local  variable, to be
called $k_t$, which is defined along the directrix so that we have
\begin{eqnarray}
\label{lockt} 
k_{t
j}^2=\frac{(2k_{j-1}k_j)(2k_jk_{j+1})}{(2k_{j-1}k_j)+(2k_jk_{j+1})
+(2k_{j-1}k_{j+1})}\equiv \frac{s_{j-1,j} s_{j, j+1}}{s_{j-1,j,j+1}}
\end{eqnarray}
in  terms of  the  energy--momentum vectors  of  the (colour)  adjacent
gluons.

We note that the two values  of $k_{\perp}$ and $k_t$ coincide for the
last emitted gluon. But in  general the transverse momentum $k_t$ of a
gluon may be  much smaller than the $k_{\perp}$ value  at which it was
emitted, because  of the recoils from subsequent  emissions.  In fact,
in this way $k_t$ can even become smaller than the resolution scale at
which the state is being sampled. But the ``emission'' $k_{\perp}$ and
the concepts of the ``first emitted'' gluon and the ``second emitted''
gluon  etc are  necessities of  the strategies  used to  simulate this
quantum mechanical process,  and not properties of the  final state of
partons. When one asks what the probability is for the production of a
state with a certain number of partons, where colour connected partons
with  transverse momenta  smaller than  a certain  value would  not be
considered  as  resolved   emissions,  the  only  relevant  transverse
momentum is the $k_t$, which  is calculated using vectors in the final
state.  Therefore, it  is undesirable that this variable  is pushed to
values  smaller  than  the  resolution  parameter. We  have  used  two
different methods to avoid such situations.  The first is to veto such
emissions  along  the   cascade\footnote{This  feature  will  be  made
available in the  next version of {\Ar}} and the  second to change the
ensuing directrix vectors so that three neighbouring lightlike vectors
are  combined into  two.  All  the results  of this  investigation are
independent of the method that is used.  Furthermore, such a procedure
is  essential for  the inclusion  of the  second order  matrix element
corrections into {\Ar} \cite{LL2}.

\subsection{The $\lambda$-measure, its properties and its connection
with the $\ell$-measure}

When a  string without gluonic  excitations fragments under  the usual
Lund Model assumptions,  hadrons are formed by breaking  of the string
field  at  ``vertices''  which,  on  the average,  lie  on  a  typical
hyperbola  parametrised by  a  squared proper  time  from the  origin,
usually  called $\Gamma_0$  or $m_0^2$.  The rapidity  range available
along this hyperbola is given by $\ln\frac{s}{m_0^2}$.

In  the Lund Model  interpretation for  a string  with a  single gluon
excitation, there  will be two parts  of the string;  one string piece
spanned between the  ${\q}$ and the ${\g}$ and  one between the ${\g}$
and  the ${\qbar}$.  Therefore  there will  be two  hyperbolic angular
ranges $(\Delta  y)_{01}$ and $(\Delta y)_{12}$  for hadron formation.
The sum of these ranges is
\begin{eqnarray}
\label{2hadronphase}
(\Delta              y)_{01}+(\Delta             y)_{12}             =
\ln(\frac{s_{01}}{2m_{0}^2})+\ln(\frac{s_{12}}{2m_{0}^2})=
\ln(\frac{s}{m_{0}^2})+\ln\left(\frac{s_{01}s_{12}}{4s m_{0}^2}\right)
\end{eqnarray}  

We draw attention to the  factor of 4 in Eq. (\ref{2hadronphase}), and
its absence in  Eq. (\ref{2gluonphase}).  In the Lund  Model, when the
partons move  apart stretching a string-like field  between them, they
lose energy--momentum to the string  field. A gluon is attached to two
string  pieces  while  a  quark   or  an  anti-quark  is  attached  to
one. Therefore a gluon will  lose energy--momentum twice as quickly as
a  quark or  an  anti-quark.  Flat string  regions  will therefore  be
formed, bounded by the whole  of quark or anti-quark momenta, but half
of gluon momenta.  Since hadrons form from this field, the phase space
is calculated by  adding up the lengths of  typical hyperbolae in each
of these flat string regions.  For instance, the invariant mass of the
region  involving the  quark  and the  gluon,  for the  system in  Eq.
(\ref{2hadronphase}),        is       $(k_{\mathrm{quark}}+\frac{1}{2}
k_{\mathrm{gluon}})^2=\frac{s_{01}}{2}$,  and its contribution  to the
phase space for hadronisation is $\frac{s_{01}}{2 m_0^2}$.

Such a factor was not necessary in Eq. (\ref{2gluonphase}). A gluon is
shared between two dipoles, but  both of those dipoles are modified by
the emission of  a single gluon from either of  these dipoles. This is
because the  gluon in common between  them would receive  a recoil and
that would change the invariant  masses of both the dipoles. Formation
of a  single hadron in string  fragmentation does not  give recoils to
the partons  making up the  string.  Hadron formation occurs  from the
string, which has a spatial extent. Gluon emission, on the other hand,
is a perturbative  phenomenon and the new emission  is thought to come
directly from one or the other  of the two partons making a dipole, or
from  the dipole  as a  whole regarded  as a  unit. Therefore,  in the
implementation of  the dipole cascade  model in {\Ar}, the  full gluon
momentum is used  for both the dipoles involving  the gluon.  There is
no problem in conserving energy--momentum because both the dipoles can
not emit ``simultaneously''. As soon  as there is an emission from one
dipole, the whole system is updated into a new chain of dipoles.

The expression  in Eq.  (\ref{2hadronphase})  does not make  sense for
soft or  collinear gluons, as these would  give negative contributions
to the  effective rapidity  range.  An infrared  stable generalisation
applicable for arbitrarily many  gluons was presented in \cite{BAGGBS}
and  its  relation  to  the  hadronisation process  was  discussed  in
\cite{BAPDGG2}.   For a  state  with  a single  gluon  we replace  the
expression in Eq. (\ref{2hadronphase}) with:

\begin{eqnarray}
\label{bettapprox}
 \ln\left(\frac{s}{ m_{0}^2} +\frac{ s_{01}s_{12}} {4 m_{0}^4}\right)
\end{eqnarray}
This is a  nicely interpolating expression between the  results for a
string  with  and  without a  gluon.   It  can  be generalised  to  an
arbitrary number  of gluons  by introducing the  following quantities,
defined  as integrals  along the  directrix $\mathcal{  A}(\xi)$ (with
$\xi$ a suitable parameter, here taken to be the energy along the directrix):
\begin{eqnarray}
\label{Adividen}
t_j    &=&    \int_0^{E_{tot}}\!\!    d\xi_1\int_0^{\xi_1}d\xi_2
\left[\frac{d\mathcal{ A}}{d\xi_1} \cdot \frac{d\mathcal{ A}}{d\xi_2}\right]
\int_0^{\xi_2}\!\!  d\xi_3\int_0^{\xi_3}d\xi_4  \left[\frac{d\mathcal{
A}}{d\xi_3}   \cdot   \frac{d\mathcal{  A}}{d\xi_4}\right]
    \cdots     \\    \nonumber    &     &\cdots    
\int_0^{\xi_{2j-2}}\!\!      d\xi_{2j-1}\int_0^{\xi_{2j-1}}
d\xi_{2j}      \left[      \frac{d\mathcal{A}}{d\xi_{2j-1}}            \cdot
\frac{d\mathcal{A}}{d\xi_{2j}}\right]
\end{eqnarray}
This  corresponds to a  $2j$-fold partitioning  of the  directrix into
non-overlapping  pieces and  then to  multiplying the  adjacent vector
differentials two by two. For  a massless relativistic string, a point
on the string (parametrised by the energy ($\sigma$) between the point
and the quark end) at time $t$ will be at $x(\sigma,t)$:
\begin{eqnarray}
\label{stringpoint}
x(\sigma,t)=\frac{[\mathcal{ A}(t+\sigma)+\mathcal{ A}(t-\sigma)]}{2}
\end{eqnarray}
i.e.   it is  given by  one  left-moving and  one right-moving  vector
defined  by points  on  the  directrix. As  the  differentials of  the
directrix  are  lightlike  the  quantities $d\mathcal{  A}(\xi)  \cdot
d\mathcal{  A}(\xi^{\prime})$  occurring   in  the  integrals  in  Eq.
(\ref{Adividen}) are surface elements on the string.  The integrals in
this  way correspond to  all the  possibilities of  obtaining $j$-fold
partitionings into non-overlapping areas on the string surface.

We may then define a functional $T \equiv \exp(\lambda)$ as
\begin{eqnarray}
\label{Tdef}
T = 1 + \sum_{j=1}^{\infty} \frac{t_j}{(m_0^2)^j}
\end{eqnarray}
i.e. as the generating  function of the partitioning functionals $t_j$
in  Eq.   (\ref{Adividen}) with  $m_0$  an  energy  scale to  make  it
dimensionless.  We note  that the integrals $t_j$ will  vanish if $j >
n+1$ where $n$ is the number of gluons emitted. The term involving one
power  of $m_0$  is always  equal to  $\frac{s}{2m_0^2}$ and  the last
(nonvanishing) term will have the generic form
\begin{eqnarray}
\label{lastterm}
\frac{1}{2}\frac{s_{01}}{2m_0^2}\frac{s_{12}}{4m_0^2}\frac{s_{23}}{4m_0^2}\cdots
\frac{s_{n,n+1}}{2m_0^2}
\end{eqnarray}
with $(j,j+1)$ indices of colour-consecutive partons  ( index $0$
for the ${\q}$, indices 1 .. n for gluons, and $n+1$ for the ${\qbar})$.
Besides the addition of $1$ and  a factor of $1/2$ (both introduced to
obtain simple factorisation properties, cf.  below) the $T$ functional
in Eq.  (\ref{Tdef})  coincides with the argument of  the logarithm in
Eq. (\ref{bettapprox}).

It is  useful to define  a generalisation, $T(\xi)$, by  exchanging the
upper  limit   in  Eq.   (\ref{Adividen})  from  $E_{tot}$   to  $\xi$
\cite{BAGGBS}. This functional fulfils
\begin{eqnarray}
\label{Tint1}
T(\xi)=1+\int_0^{\xi}\int_0^{\xi_1}\frac{   d\mathcal{  A}(\xi_1)\cdot
d\mathcal{ A}(\xi_2)}{m_0^2} T(\xi_2)
\end{eqnarray}
and together with the four-vector valued functional $q_T$\footnote{Called $q$ in \cite{BAGGBS}}:
\begin{eqnarray}
\label{qTdef}
q_T(\xi)=        \frac{1}{T(\xi)}        \int_0^{\xi}       d\mathcal{
A}(\xi^{\prime})T(\xi^{\prime})
\end{eqnarray}
we obtain the differential equations
\begin{eqnarray}
\label{diffeq}
dT   &=&   \frac{q_Td\mathcal{   A}}{m_0^2}   T  \nonumber   \\   dq_T
&=&d\mathcal{  A}  -  \frac{q_T d\mathcal{  A}}{m_0^2}q_T  \nonumber\\
dq_T^2&=& 2(1-\frac{q_T^2}{m_0^2})q_Td\mathcal{ A}
\end{eqnarray}
They can be integrated to
\begin{eqnarray}
\label{Tint2}
T(\xi)&=&\exp\left(\int_0^{\xi}\frac{q_T(\xi^{\prime})d\mathcal{
      A}(\xi^{\prime})}{m_0^2}\right)  \nonumber   \\  q_T^2(\xi)  &=&
      m_0^2[1-T(\xi)^{-2}]
\end{eqnarray}

Finally,   we  define   a  four-vector   valued   function  $\mathcal{
X}_{\mu}=\mathcal{ A}_{\mu}-{q_{T}}_{\mu}$\footnote{Called $X$ in \cite{BAGGBS}} so that
\begin{eqnarray}
\label{calXdef}
d\mathcal{ X}=\frac{(q_Td\mathcal{ A})}{m_0^2}q_T
\end{eqnarray}
i.e.   the vector  $q_T$ is  the tangent  to the  $\mathcal{ X}$-curve
reaching out to the directrix. From Eq. (\ref{Tint2}) we conclude that
the functional $T$ is the exponential of an area (note that the vector
$d\mathcal{  A}$ is  lightlike, and  $q_{T}d\mathcal{ A}$  is  an area
element  between the  directrix and  the $\mathcal{  X}$-curve).  This
area  is proportional  to the  functional $\lambda$  as mentioned  in the
Introduction. We also  note that the vector $q_T$  quickly reaches the
length $m_0$.  Integrating the  differential equations for a directrix
built  up from  a set  of lightlike  vectors $\{k_j\}$  we  obtain the
iterative equations \cite{BAGGBS}:
\begin{eqnarray}
\label{Tint3}
T_{j+1}&=&           (1+q_{T       j}k_{j+1}/m_0^2)    T_j      \equiv
\frac{T_j}{\gamma_{j+1}} \nonumber \\ q_{T j+1}&=& \gamma_{j+1}q_{T j}
+ \frac{1+\gamma_{j+1}}{2}k_{j+1}
\end{eqnarray}
The starting values are $T_0=1$ and $q_{T 0}=k_0$.

The factorisability  of the functional  $T$ means that  its logarithm,
 $\lambda$,  can be  written as  a sum  containing one  term  for each
 $k_j$:
\begin{eqnarray}
\label{lambdapart}
\lambda  \equiv \sum_{j=1}^{n+1}  \Delta \lambda_j  = \sum_{j=1}^{n+1}
\ln\left(1+q_{T j-1}k_{j}/m_0^2\right)
\end{eqnarray}
\begin{figure}
\begin{center}
\epsfig{figure=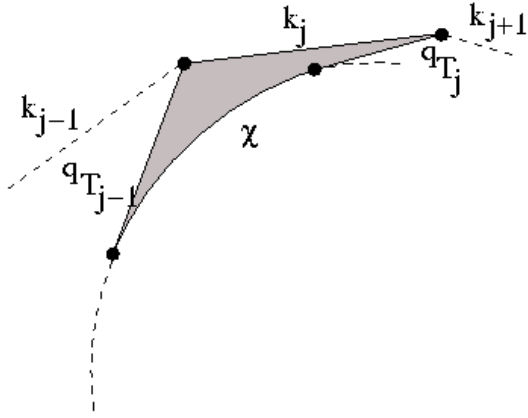, width=7.0cm }
\caption{A plaquette bordered by $q_{T  j}$, $q_{T j+1}$ , $k_j$ and a
hyperbolic segment from the  $\mathcal{ X}$-curve, as described in the
text.}
\label{deltalambdaarea}
\end{center}
\end{figure}
The $\Delta  \lambda_j$ defined in this  way is the size  of a subarea
just as  the total $\lambda$  represents the area spanned  between the
$\mathcal{     X}$      and     the     directrix      curves.      In
Fig. \ref{deltalambdaarea}  we exhibit such  a region bordered  by the
``initial''  $q_{T j-1}$ and  the ``final''  $q_{Tj}$ together  with a
hyperbolic  segment  from  the  $\mathcal{  X}$-curve  and  the  gluon
energy--momentum vector $k_j$ from the directrix.

If     we     define     $\gamma_0\equiv1$     and     the     vectors
$\acute{k_j}\equiv\frac{1}{2}k_j(1+\gamma_j)$, then it  is easy to see
that  the vector  $q_{T j}$  is a  weighted sum  of  the $\acute{k_l}$
vectors ($l  \leq j$).  The weight for  any vector  $\acute{k_{m}}$ is
exponentially suppressed with $\sum_{l=m}^{j-1} \Delta \lambda_{l+1}$.

To see  that $\Delta \lambda_j$  as defined in  Eq. (\ref{lambdapart})
actually has the meaning of a rapidity  we go to the rest frame of the
vector $q_{T  j-1}$ and assume for  simplicity that it  has the length
$m_0$.   Further assume that  the lightlike  vector $k_j$  is directed
along the $3$-axis with the  length $\omega$ so that $\Delta \lambda_j
=  \ln(1+\omega/m_0)$. Then  it  is  easy to  show  that the  ``next''
$q_T$-vector $q_{T  j}$ will have the lightcone  coordinates along the
$3$-axis   equal  to   $(m_0\exp(\Delta   \lambda_j),  m_0\exp(-\Delta
\lambda_j))$, i.e. the $q_T$ vector  has been accelerated from rest to
the  rapidity  $\Delta \lambda_j$  by  ``the  step''  $k_j$ along  the
directrix. The corresponding segment  of the $\mathcal{ X}$-curve is a
part of a hyperbola.

For sufficiently  well ordered emissions  and a resolution  scale much
smaller than  the smallest transverse momentum present,  the last term
(Eq. (\ref{lastterm})) will dominate in Eq. (\ref{Tdef}) and therefore
$\lambda(m_0=\frac{1}{2}k_{\perp})$    will    be   approximately equal to 
$\ell(k_{\perp})$ defined in Eq. (\ref{ngluonphase}).

From  the definition  in Eq.   (\ref{lambdapart}) it  is easy  to show
(cf. the definition of the vectors $q_{Tj}$ in Eq. (\ref{Tint3})) that
the argument  in the logarithm of  the $\Delta \lambda_j$  is equal to
$(1+s_{j-1,j}/4m_0^2)$  plus  terms  down  by  at least  a  factor  of
$\gamma$.   From   the  definition  of  the   local  $k_t$  variables,
Eq. (\ref{lockt}), we may also conclude that
\begin{eqnarray}
\label{ktsize}
k_{tj}^2 < \frac{s_{j-1,j}s_{j,j+1}}{(s_{j-1,j}+s_{j,j+1})}
\end{eqnarray}
and consequently we obtain
\begin{eqnarray}
\label{sizektlambda}
\Delta                          \lambda_j                          &>&
\ln(\frac{s_{j-1,j}}{4m_0^2})=\ln(\frac{s_{j-1,j}}{k_{\perp}^2})
\equiv \ell_{j-1,j} \nonumber \\ \ln(\frac{k_{tj}^2}{k_{\perp}^2}) &<&
\min(\ell_{j-1,j},\ell_{j,j+1})
\end{eqnarray}
This  implies  that there  is  a  minimum  positive value  of  $\Delta
\lambda$, if a  procedure to exclude or filter  out emissions creating
gluons with $k_t<k_{\perp}$ has been used.
\section{The Properties of the $\Delta \lambda$ Distributions}
\label{results}

\begin{figure}
\begin{center}
\rotatebox{270}{\epsfig{figure=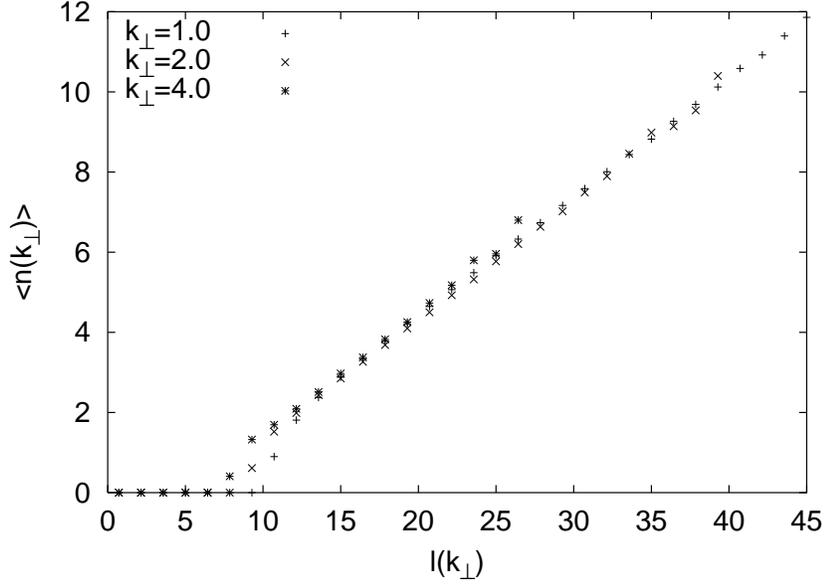,
width=0.5\textwidth }}
\caption{The   figure  shows   a  plot   of   $\langle n(k_{\perp})\rangle$  versus
$\ell(k_{\perp})$ for an $e^+e^-$-event  with $\sqrt{s}=$ 200 GeV from
{\Ar}.}
\label{nversuslambda}
\end{center}
\end{figure}

We  will  now exhibit  a  very  noticeable  regularity of  the  dipole
cascades, cf. Fig. \ref{nversuslambda}. In this figure, we show a plot
of  the  average  number  of  gluons as  a  function  of  $k_{\perp}$,
$n(k_{\perp})$,  versus  $\ell(k_{\perp})$  for  $e^+e^-$-events  with
$\sqrt{s}=200$ GeV and  for a range of values  of $k_{\perp}$ from $4$
down to $1$  GeV. We have switched off the  gluon splitting process so
that we end up with one colour connected set of partons.
 
\begin{figure}
\begin{center}
\rotatebox{270}{\epsfig{figure=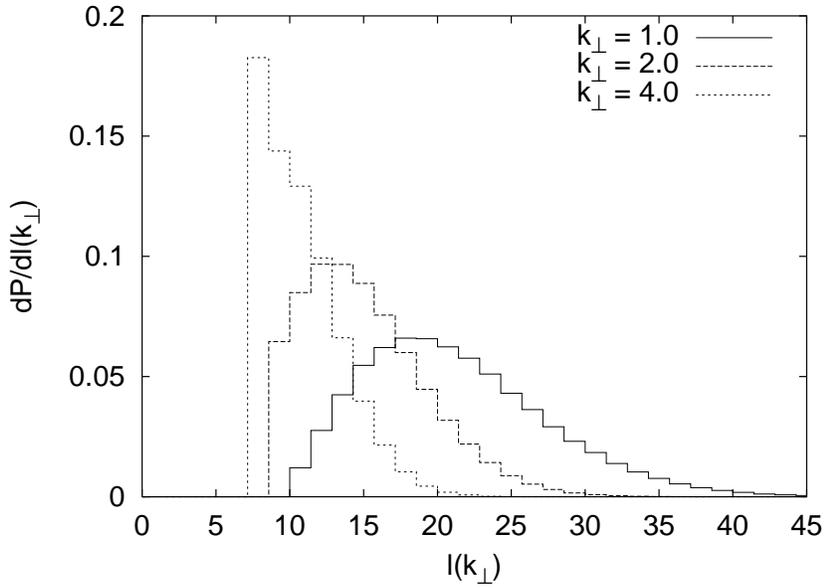,
width=0.5\textwidth }}
\caption{The      $\ell(k_{\perp})$-distribution     for     different
 $k_\perp$-values at $\sqrt{s}=$ 200 GeV.}
\label{lambdakt}
\end{center}
\end{figure}

\begin{figure} 
\begin{center}
\epsfig{file=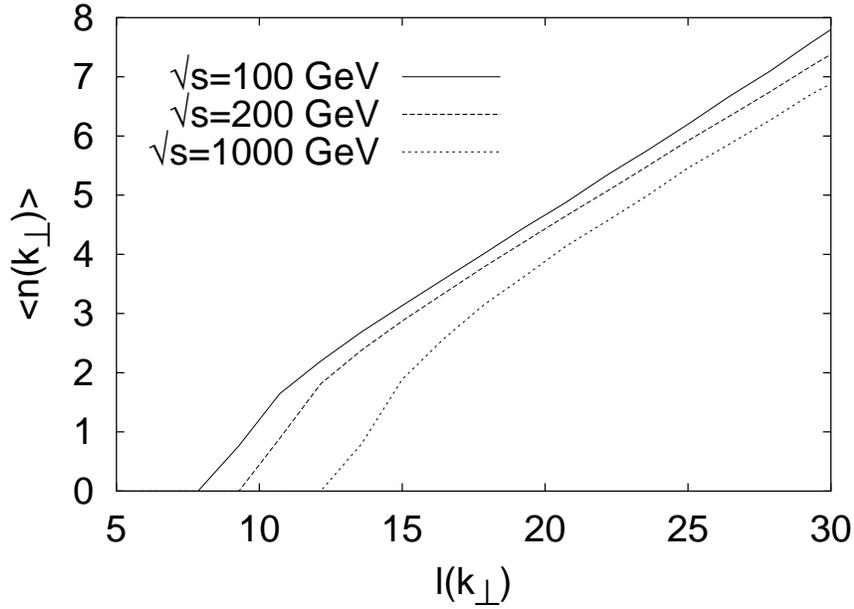,width=12.0cm}
\caption{The   figure  shows   a  plot   of   $\langle n(k_{\perp}) \rangle$  versus
$\ell(k_{\perp})$ ($k_{\perp}=1.0$ GeV) for different cms energies from {\Ar}. }
\label{nversuslambdaenergy}
\end{center}
\end{figure}

There are several interesting properties of this plot:

\begin{itemize}

\item  There  is  a  linear  correlation  between  the  average  gluon
multiplicity  $\langle  n(k_{\perp})  \rangle$  and  the  phase  space
variable $\ell(k_{\perp})$  inside a very narrow band.   The fact that
the different  lines obtained for different values  of $k_{\perp}$ are
so close to each other means that the slope of this linear correlation
is  rather insensitive  to changes  in the  ordering variable  in this
range.   It should be  noted that  the $\ell(k_{\perp})$-distributions
are widely  varying for  these $k_{\perp}$-values as  we show  in Fig.
\ref{lambdakt}.  We have used  a running coupling (with ${\LQCD}=0.22$
GeV,   i.e.   the   default  value   in  {\Ar}).    The   region  with
$\ell(k_{\perp}) > 20$  is mainly populated by the  results from small
$k_{\perp}$-values, $k_{\perp}<2$ GeV.

\item We find that the linear correlation in Fig.  \ref{nversuslambda}
is independent of the cms  energy (we have used different cms energies
ranging from  $40$ GeV  and up  to even $2000$  GeV, though  for small
energies there are too few  gluons emitted to allow detailed studies),
cf.  Fig.   \ref{nversuslambdaenergy}.  Not only  do we find  a linear
correlation at each value of cms  energy (when the number of gluons is
more than 2),  the slope of the correlation  varies rather weakly with
respect to energy.  This property is  also not a function of the total
thrust, or the presence of particularly hard emissions (e.g.  gluon of
$k_{\perp}>7$ GeV for $\sqrt{s}=200$ GeV). There is a minimum value of
$\ell \simeq  10$, corresponding to no gluon  emission (for $k_{\perp}
=1$ GeV  and $\sqrt{s}=200$  GeV), and there  is a  (small) transition
region for a  few more units, until the number  of gluons exceeds two.
But after that $dn/d\ell$ is a constant.

\item As a consistency check  we have used two procedures for sampling
the events. It is possible to  run the parton cascade in {\Ar} down to
a certain value  of the ordering variable $k_{\perp}$,  i.e.  emit all
gluons with  larger $k_{\perp}$'s,  sample the emerging  directrix and
investigate  its  properties, e.g.   calculate  the  number of  gluons
emitted  to  that  point,  $n(k_{\perp})$,  the total  length  of  the
$\mathcal{   X}$-curve,  $\ell(k_{\perp})$,  the   individual  $\Delta
\lambda(m_0)$'s and the local $k_t$'s.  After that we may continue the
cascade for the  same event down to smaller  $k_{\perp}$-values and do
the same exercise.  Another possibility  is to run the cascade down to
a  certain  value $k_{\perp}\equiv  {k_{\perp}}_{cut}$  and collect  a
large sample of  events to investigate.  After this we  can pick a new
value of  ${k_{\perp}}_{cut}$ and repeat  the process.  The  result in
Fig. \ref{nversuslambda} does not depend upon which of the two methods
for sampling the events we use.
\end{itemize}
\begin{figure} 
\begin{center}
\mbox{
\hspace{-0.5cm}
\rotatebox{270}{\epsfig{file=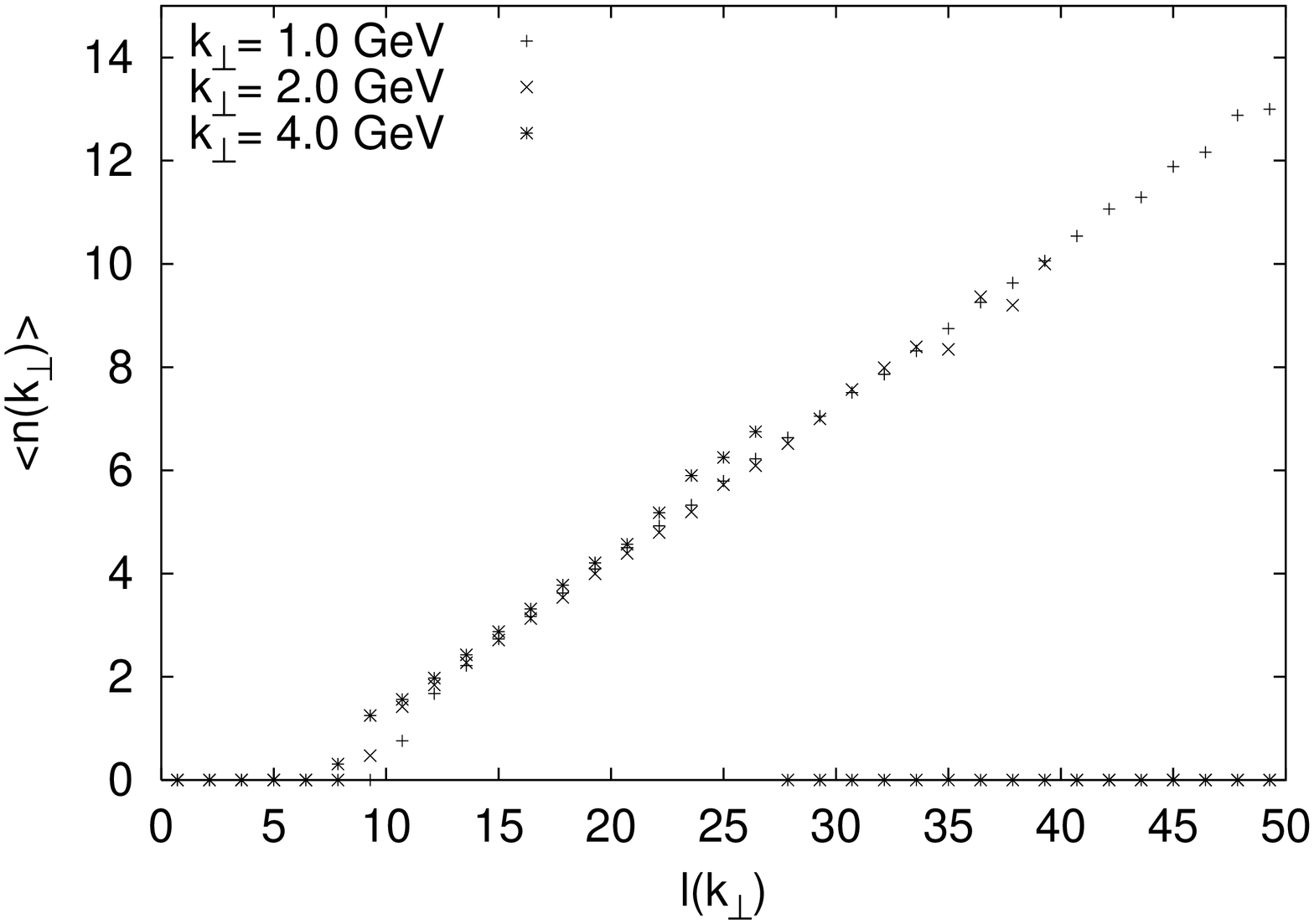,width=6.0cm}}
\hspace{-0.5cm}
\rotatebox{270}{\epsfig{file=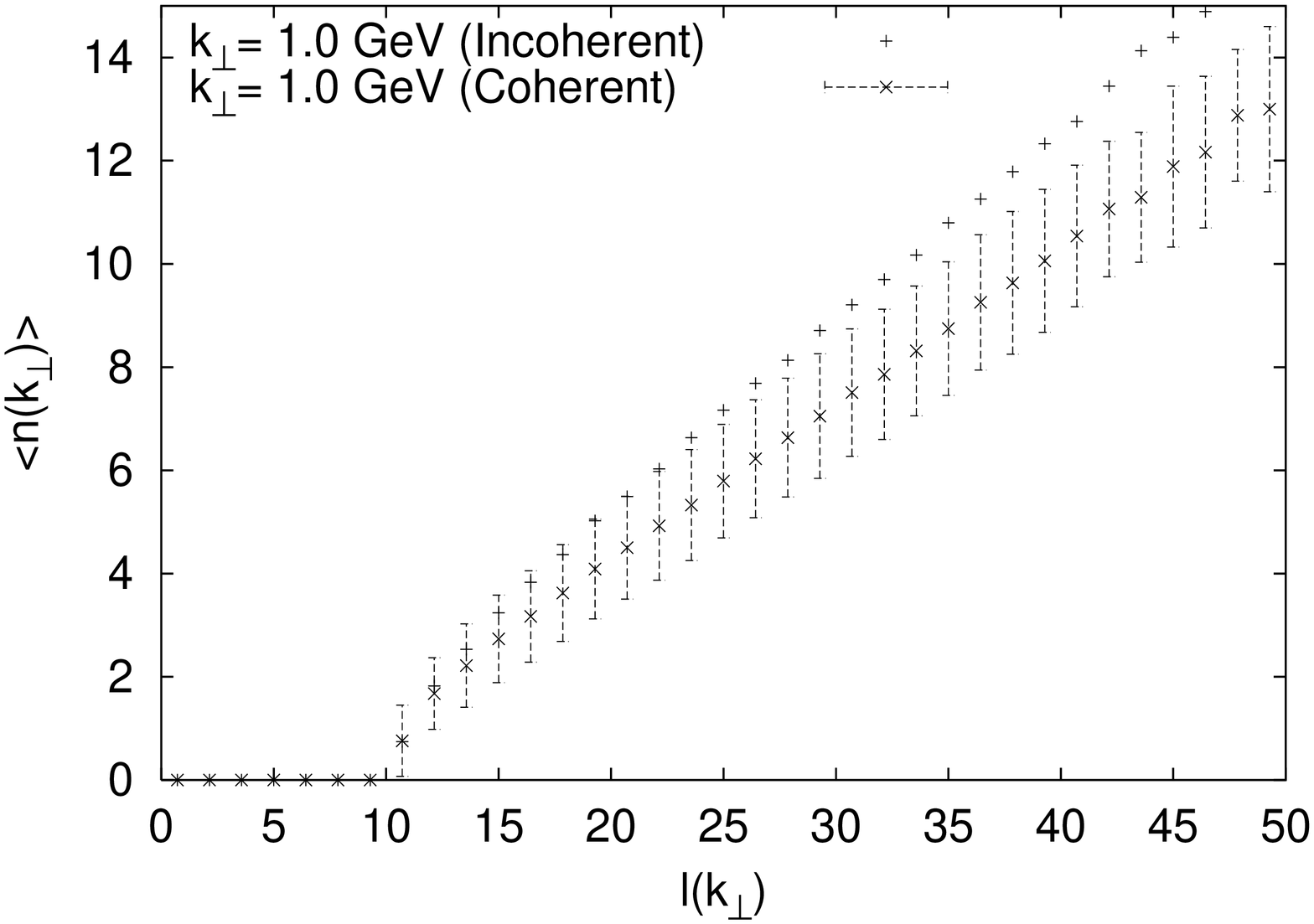,width=6.0cm}}       }
\captive{The  figure to  the  left shows  a  plot of  $\langle n(k_{\perp}) \rangle$
versus  $\ell(k_{\perp})$ for an  $e^+e^-$-event with  $\sqrt{s}=$ 200
GeV from {\Py}.  The figure to the right  shows $ \langle n(k_{\perp}) \rangle$ $\pm$
$\sigma$ versus $\ell(k_{\perp})$ at $\sqrt{s}=$ 200 GeV, $k_{\perp}=$
1.0 GeV for coherent and incoherent showers in {\Py}.
\label{jetsetnla}
}
\end{center}
\end{figure}

We  observe the  same  effect in  the  parton shower  in  {\Py} as  in
{\Ar}. The ordering  variable in {\Py} is the  parton virtuality $Q^2$,
and  the variable $\hat{k}_{\perp}^2  = z(1-z)Q^2$  (where $z$  is the
lightcone  fraction  taken by  the  emitted  parton)  is used  as  the
argument for the  running coupling.  This is a  good approximation for
the   transverse   momentum   in   the   case   of   massless   parton
kinematics.  However,  since transverse  momentum  itself  is not  the
ordering  variable in  this  case, there  is  no real  lower limit  to
it. Therefore, in order to make the comparisons we filtered the events
obtained  from {\Py}  and combined  partons such  that all  gluons had
$k_t$ above a certain value proportional to $Q$.

It is  then possible  to use  {\Py} to obtain  an inclusive  sample of
$n_g(CQ)$ and  $\ell(C Q)$. We just  note that it is  possible to tune
the  constant $C$ such  that, as  shown in  Fig.  \ref{jetsetnla},
there is the same $dn/d\ell$ as in {\Ar}. We find that $C \approx 0.5$  

It is  possible to switch off  the coherence conditions  in the parton
shower in {\Py}, since the strong angular ordering is implemented only
after the  generation using the  ordering variable $Q^2$. We  show the
effect of this  feature in Fig.  \ref{jetsetnla}. For  the same $C$ we
obtain  many more  partons  per  unit phase  space  for an  incoherent
shower,  although  they  still   show  a  linear  correlation  to  the
$\ell$-measure.

\begin{figure}[t]
\begin{center}
\rotatebox{270}{\epsfig{figure=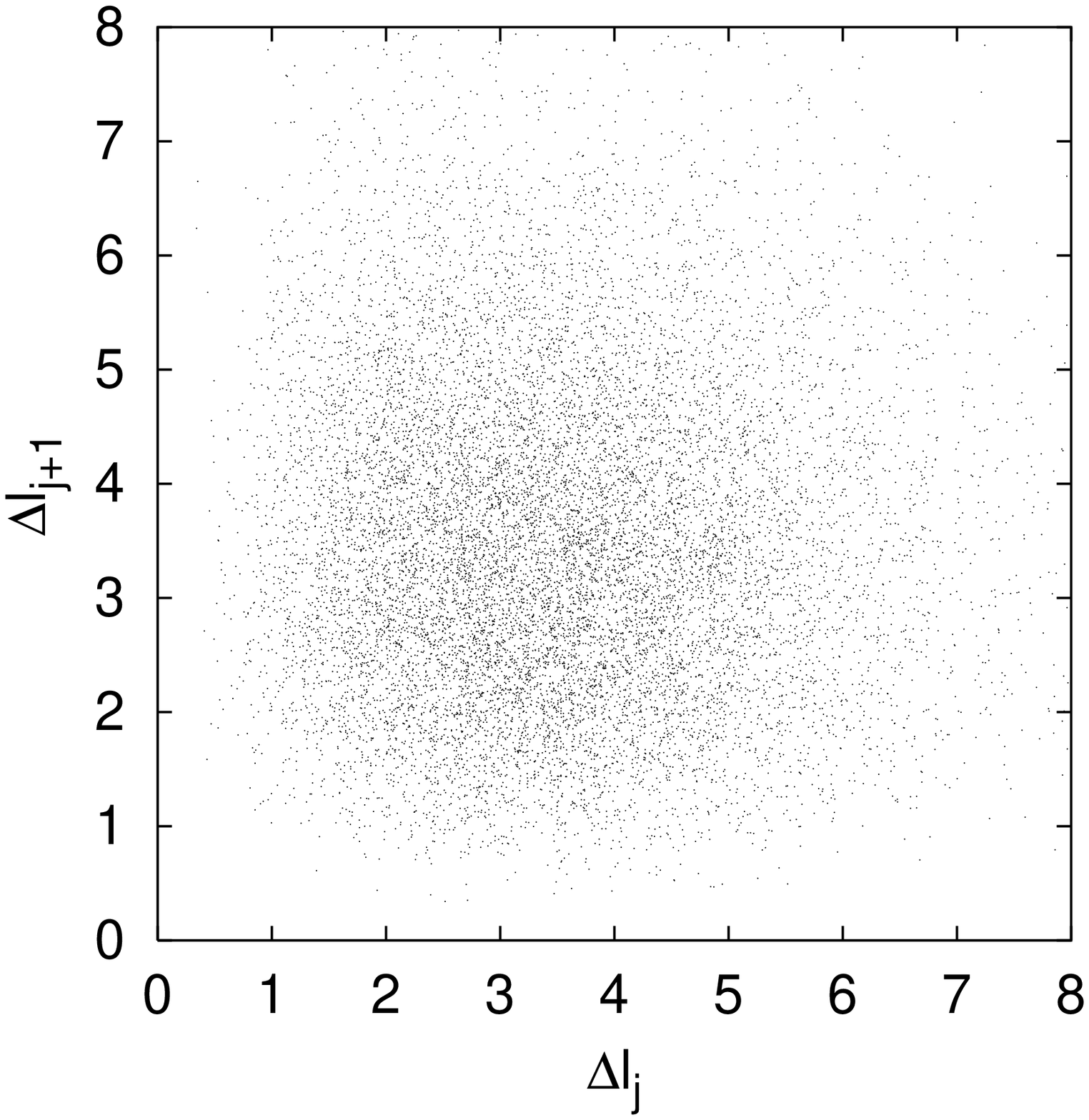,
width=0.5\textwidth }}
\caption{Scatter plot of two adjacent $\Delta\ell$ values}
\label{corrDL}
\end{center}
\end{figure}

The  linear relation  between $\langle  n \rangle$  and $\ell$  can be
understood  as follows.  If one  considers a  variable such  as $\ell$
defined as in Eq.  (\ref{ngluonphase}) where the parts $\Delta \ell_j$
are  distributed   with  the   averages  $\mu_j$  and   the  variances
$\sigma^2_j$,  then   $\ell$  itself   will  mostly  have   a  gaussian
distribution according  to the central limit theorem.   The average is
$\bar{\mu}  =  \sum  \mu_j  $ and  the  variance  $\bar{\sigma^2}=\sum
\sigma^2_j  $.  The  numbers $\bar{\mu}$  and $\bar{\sigma^2}$  are in
general well-defined  for large $n$  unless the values of  the $\Delta
\ell_j$  are correlated.  We  find that  the $\Delta  \ell_j$'s indeed
show very little correlation,  even between adjacent dipoles, cf.  Fig
\ref{corrDL}. We also find that  the average $\mu_j$ shows only a weak
dependence upon the total number of gluons in the event, n, for a wide
range of n.
\begin{figure}[t]
\begin{center}
\rotatebox{270}{\epsfig{figure=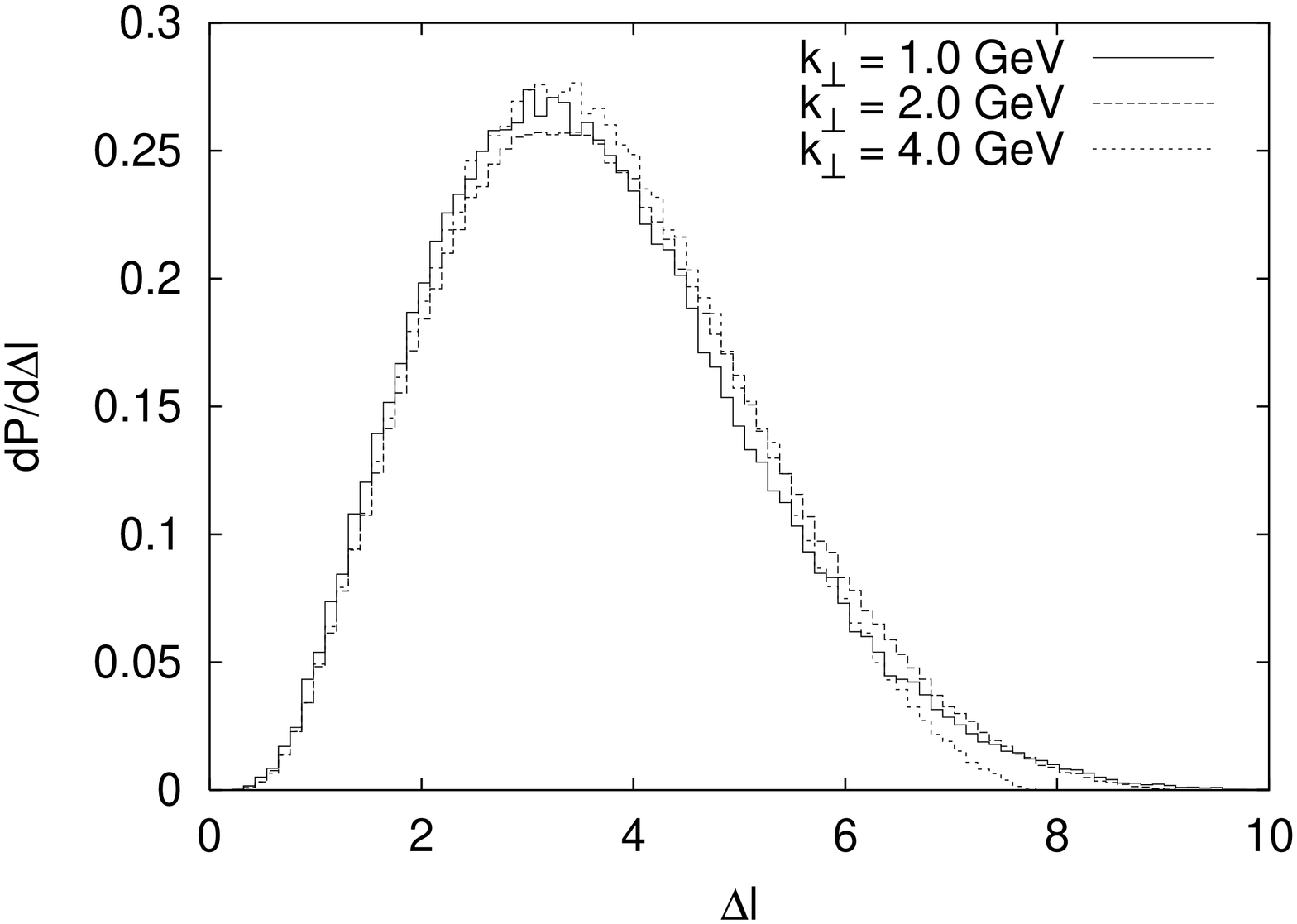,
width=0.5\textwidth }}
\caption{The   $\frac{dP}{d(\Delta   \ell)}   \equiv   P_{\ell}(\Delta
\ell)$-distribution from {\Ar}.}
\label{Deltalambda}
\end{center}
\end{figure}

We will therefore go over  to the investigation of the distribution of
the $\Delta  \ell_j$ defined  in Eq.  (\ref{ngluonphase}).   We notice
that  the first  and the  last  term in  Eq.  (\ref{ngluonphase})  are
different  from the  other terms  in  one important  aspect. They  are
contributions  from dipoles involving  a quark  or an  antiquark (with
different  colour charges than  gluons), whereas  all the  other terms
come from purely gluonic dipoles. It turns out that the dipoles at the
ends  have  slightly  different   distributions  than  all  the  rest.
However, if the gluon splitting process is switched off, there are only
two such  dipoles in an  event. This means  that we can write  for the
average    $\bar{\mu}   =    2 \langle \mu_{{\q}} \rangle+(n-1) \langle 
\mu_{\mathrm{gg}} \rangle$.     In   Fig.
\ref{nversuslambdaenergy} we  see this  effect clearly, because  it is
only for $n>2$ that we find a constant slope, i.e. when we have purely
gluonic  dipoles present.  The  properties we  noted  above are  about
variation  of  the  number  of  gluons  or  dipoles  with  respect  to
$\ell(k_\perp)$ and  the stability of  the slope of  the correlations,
and   should  correspond   to   the  properties   of  purely   gluonic
dipoles. Therefore, from now on we will not include the very first and
the very last $\Delta \ell$-values in the investigations. We have seen
that purely gluonic dipoles,  irrespective of their position along the
directrix,   have   the   same   distribution  in   size.    In   Fig.
\ref{Deltalambda}, we show the distribution $\frac{dP}{d(\Delta \ell)}
\equiv P_{\ell}(\Delta \ell)$ from {\Ar} and observe that:
\begin{figure}[t]
\begin{center}
\rotatebox{270}{\epsfig{figure=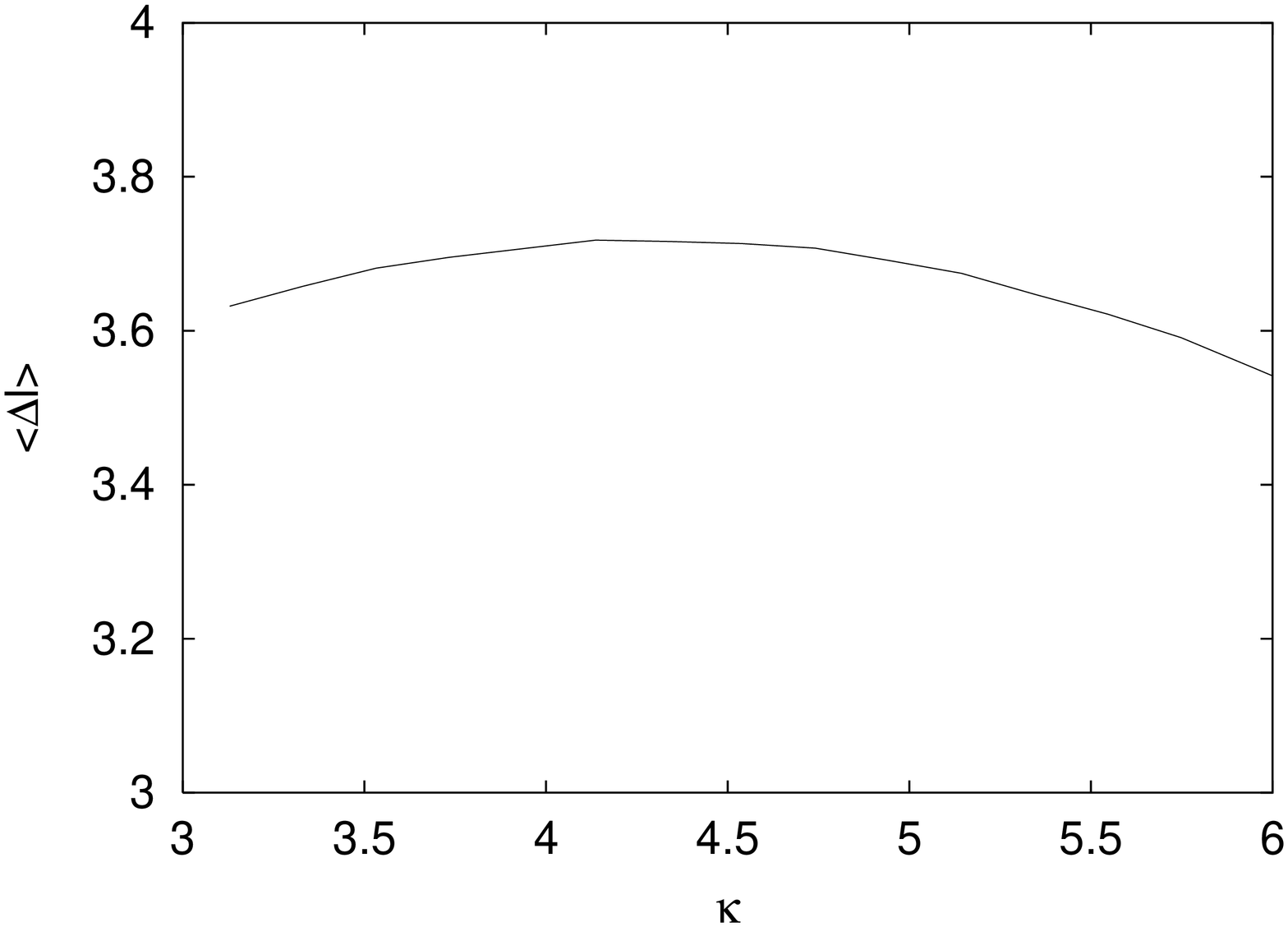,
width=0.5\textwidth }}
\caption{The average  value of $\Delta\ell$  as a function  of $\kappa
\equiv \ln(k_{\perp}^2/\Lambda^2)$. Note that the variation in the average
$\Delta \ell$ in the figure is much smaller than the width of the distribution of
$\Delta \ell$ as seen in Fig. \ref{Deltalambda}.}
\label{scaleinvariance}
\end{center}
\end{figure}

\begin{itemize}

\item The distribution $P_{\ell}$ has  an average $\mu$ and a variance
  $\sigma^2$, consistent with the results in Fig. \ref{nversuslambda}.
  We  have found  that it  is independent  of the  cms energy,  of the
  global event  variables like thrust  and the total  $\ell$-value for
  the event, and also of hard gluon emissions.

\item The  distribution $P_{\ell}$ is rather insensitive  to the value
of  $k_{\perp}$. In  Fig. \ref{scaleinvariance},  we show  the average
value   of   $\Delta  \ell$   as   a   function   of  $\kappa   \equiv
\ln(k_{\perp}^2/\Lambda^2)$ and  we find that  it changes about  5 \%,
when the ordering variable goes from $0.8$ to $3$ GeV.
\begin{figure}
\begin{center}
\rotatebox{270}{\epsfig{figure=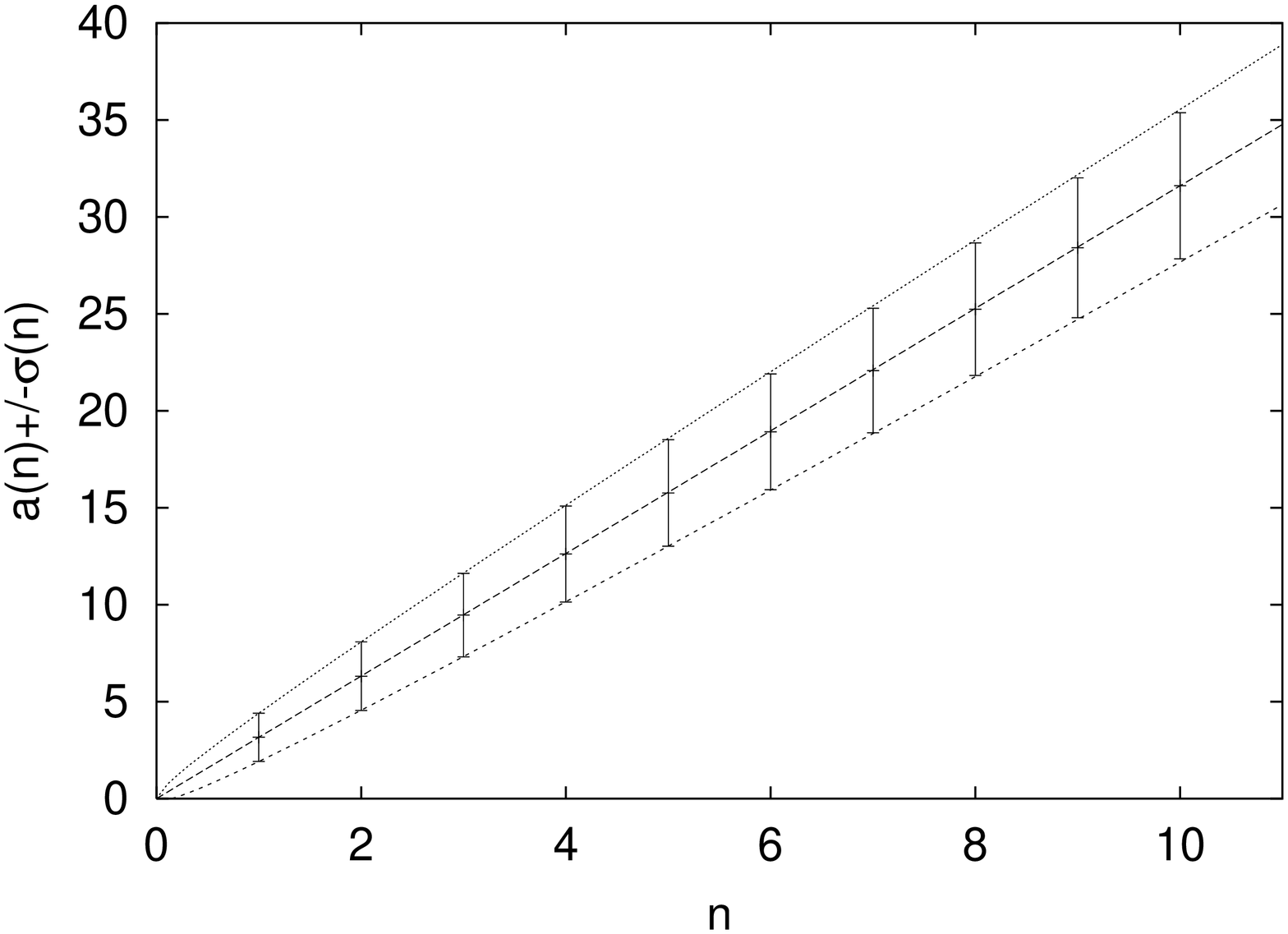,
width=0.5\textwidth }}

\caption{The figure shows the average $ \Delta \ell$ (and the
  standard deviations  around it) for chains of  $n$ connected $\Delta
  \ell$ called  $a(n)$ ($a(n)\equiv \langle \sum_{j=1}^n \Delta
  \ell_j \rangle$). The statistics  is collected from
  events with a fixed number of gluons.}
\label{P10}
\end{center}
\end{figure}
\item There  are no  noticeable correlations between  adjacent $\Delta
  \ell$'s as we have noted  before. In Fig.  \ref{corrDL}, we show the
  scatter  plot   beween  two  adjacent  $\Delta   \ell$,  taken  from
  stochastically  chosen  pairs in  many  different  events.  In  Fig.
  \ref{P10},  we show  the values  of the  average $a(n)$  (defined as
  $a(n)\equiv  \langle \sum_{j=1}^n \Delta  \ell_j \rangle$),  and the
  standard deviations  around it, for chains of  $n$ connected $\Delta
  \ell$.   For this figure  the statistics  was collected  from events
  with a fixed number of gluons. 
 
\item In order to further  investigate the independence of the $\Delta
\ell$ values  in an event we  have examined events  containing a fixed
number  ($N$) of  gluons.   We  may then  ask  about the  multiplicity
distribution,  $\hat{P}(N_{\Delta})$,   of  events  with  $N_{\Delta}$
values of $\Delta  \ell \leq \Delta$. We find  a binomial distribution
with the mean $Np$  and variance $Np(1-p)$ where $p=\int_0^{\Delta} dx
P_{\ell}(x)$  as expected  from  uncorrelated dipoles.   We have  also
investigated  the average  number of  $\Delta \ell$-values,  which are
consecutive and satisfy $\Delta \ell \leq \Delta$. We find a geometric
distribution, weighted  with the average  number of $\Delta  \ell$ not
fulfilling the condition (which is,  once again, expected when we have
uncorrelated dipoles).   We also  find that the  distribution $P_{\ell
n}$ obtained  from the sum of  the lengths of  $n$ consecutive $\Delta
\ell$ is an $n$-fold convolution of the distribution $P_{\ell}$.
 
\end{itemize}
\begin{figure}[!t] 
\begin{center}
\mbox{
\hspace{-0.5cm}
\rotatebox{270}{\epsfig{file=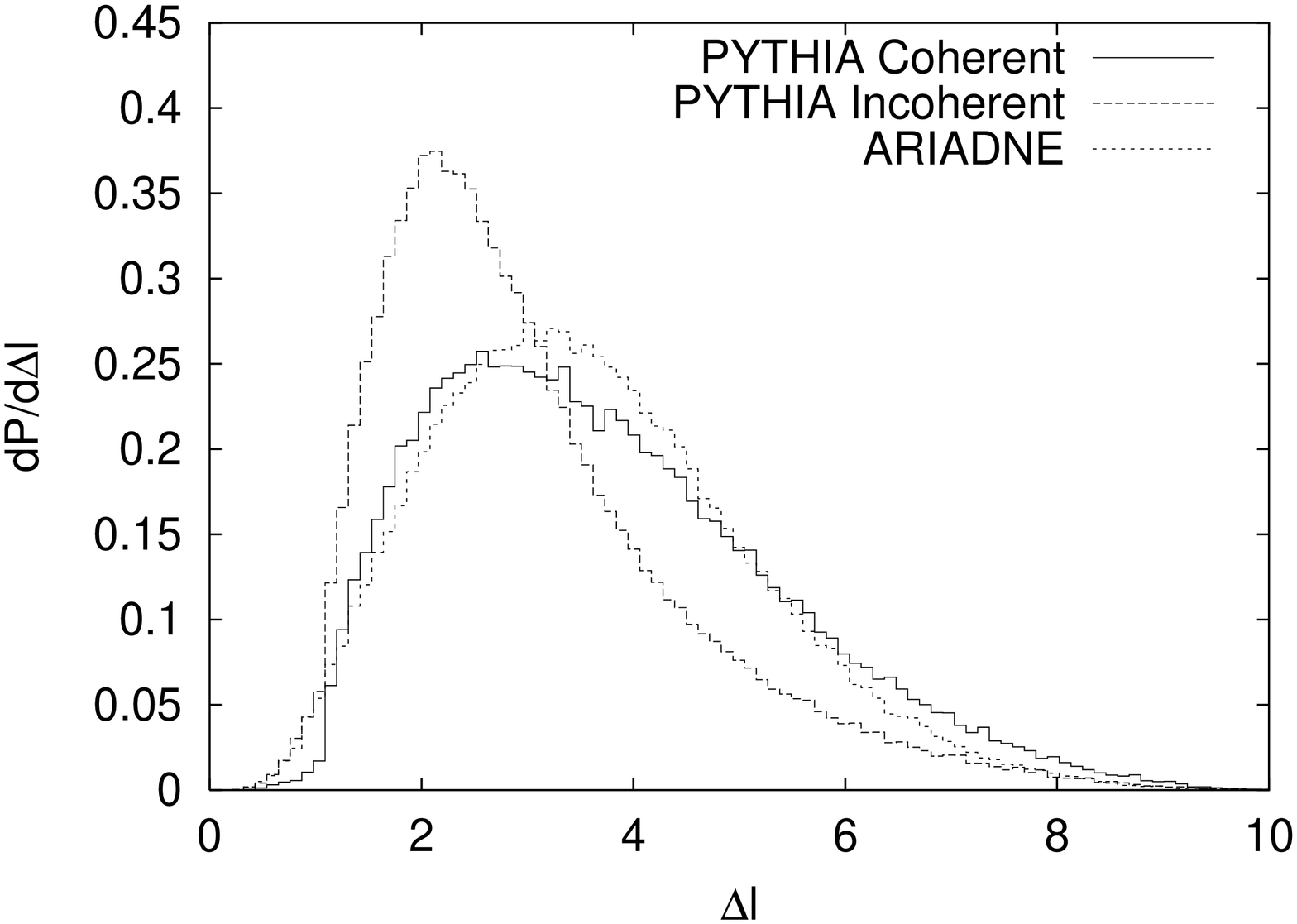,width=6.0cm}}
\hspace{-0.5cm}
\rotatebox{270}{\epsfig{file=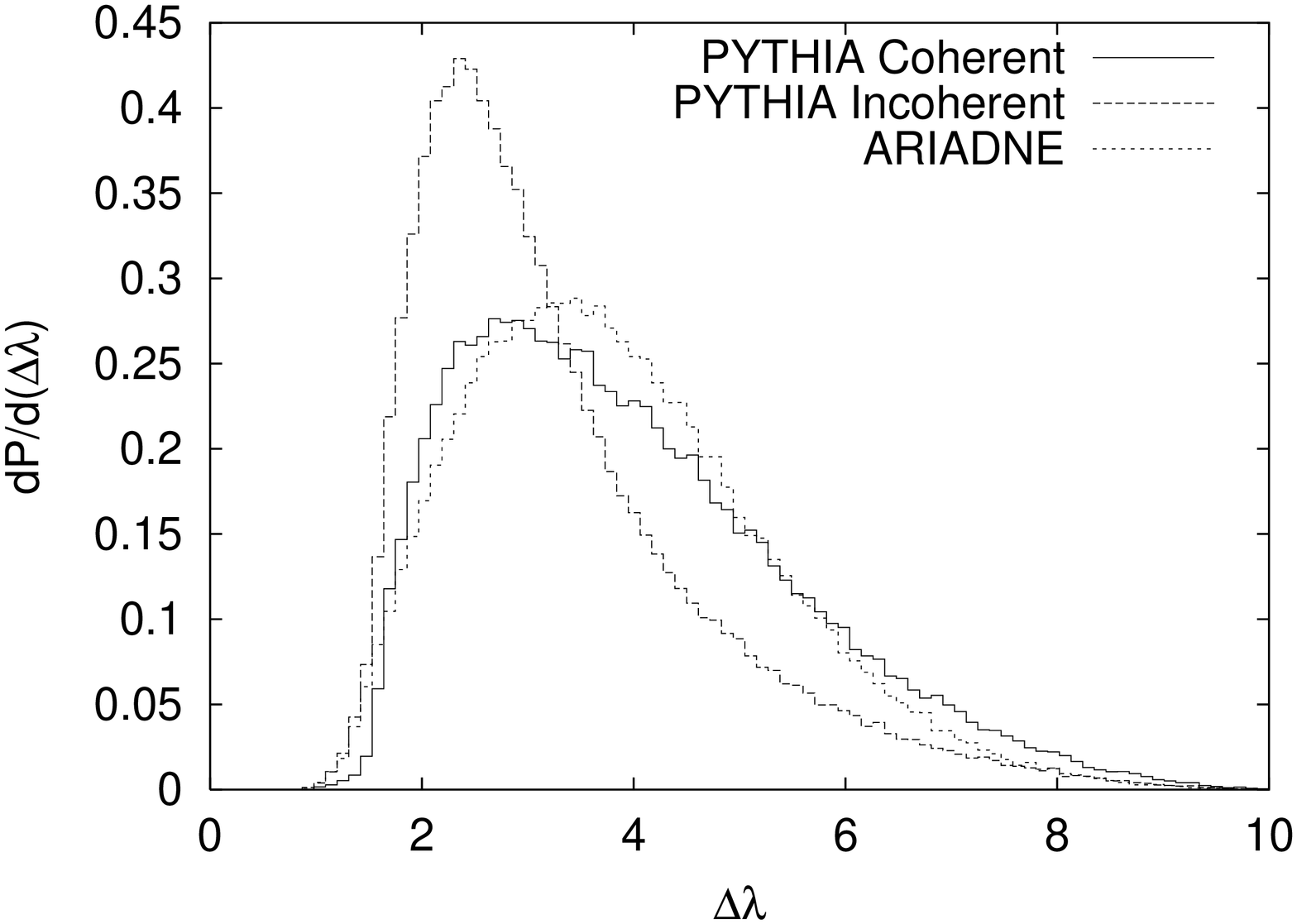,width=6.0cm}}       }
\captive{The   figure   to  the   left   shows  the   $P_{\ell}(\Delta
\ell)$-distribution from  {\Py} compared to  the distribution obtained
from {\Ar}. The  figure to the right shows  the corresponding plot for
the $P_{\lambda}(\Delta \lambda)$ distribution.
\label{jetsetDela}
}
\end{center}
\end{figure}
The corresponding results in {\Py}  are very similar to the results in
{\Ar} (with  the appropriate choice  of the constant $C$  as discussed
above).   To  the  left in   Fig.   \ref{jetsetDela}  ,   we  show  the
distributions  from  {\Py}  and  note  that  {\Py}  is  somewhat  more
concentrated  towards  small  $\Delta  \ell$  values and  also  has  a
somewhat wider  tail so  that the  average values are  the same  as in
{\Ar}.    The  ``incoherent''   option  gives   a  much   more  narrow
distribution centred around smaller values, as expected. \\

It is also  possible to investigate the distributions  in terms of the
 $\lambda$-measure   defined   in   Eq.    (\ref{lambdapart}).    In
 Fig. \ref{jetsetDela}  to the right,  we show two  distributions (one
 from  {\Ar} and  one from  {\Py}), to  be  called $P_{\lambda}(\Delta
 \lambda)$  using as  before  $2m_0=k_{\perp}=CQ$ where  $m_0$ is  the
 argument  of the  $\lambda$-measure.   Once again  we  find the  same
 properties as for the distributions $P_{\ell}(\Delta\ell)$.
 
We conclude that, inside the whole region relevant to the hadronisation
procedure in String Fragmentation, the partonic states are dominated by
this structure of independent entities. It is tempting to consider
them as a kind of collective  variables for the QCD force fields.  The
$\Delta\ell$ variables represent the local size of the dipoles between
adjacent  parton energy--momentum  vectors, whereas  the $\Delta\lambda$
values  correspond  to  a  similar although  more  non-local  property
depending upon a set of adjacent vectors.

These  properties   of  the  distributions  of   $\Delta  \ell$'s  are
properties of  gluonic dipoles, and  result from perturbative  QCD. In
the next section  we will make models in the LLA  and the MLLA schemes
based on simple gain loss  considerations on the dipole cascade model,
and show that  it is possible to understand  many qualitative features
of these distributions.

The $\Delta \lambda$'s on the  other hand are complicated objects, and
there is  no obvious  way to calculate  their distributions  except by
relying on  the similarity  between the $\lambda$-measure  and $\ell$.
But  the fact that  there are  as many  $\Delta \lambda$'s  as $\Delta
\ell$'s, and  the $\Delta  \lambda$'s have parallels  for each  of the
properties of the $\Delta \ell$'s discussed here, lead us to introduce
the notion  of Generalised  Dipoles (GD) to  be the ``source''  of the
contributions  to  the  $\Delta  \lambda_j$'s.  Whereas  a  dipole  is
spanned between two colour connected  partons, a GD corresponds to one
flat  block  or  ``plaquette''  in  the  surface  traced  between  the
directrix  and  the  $\mathcal{   X}$-curve  by  the  tangent  to  the
$\mathcal{ X}$-curve.  Regularities in the properties of dipoles which
are mirrored in the GD's would be passed on to the hadronised state in
a hadronisation scheme based on the $\lambda$-measure, such as the one
introduced in \cite{BASMFS}.

\section{Models to Describe The  Dipole Distributions}
\label{discussion}

\subsection{A Simple Cascade Model}

Multiplicity  distributions  were  presented  in  the  parton  cascade
formalism  in \cite{DFKBCMM}  and in  the dipole  cascade  formalism in
\cite{BAPDGG}  to leading  logarithmic order.   In this  paper  we will
follow the approach presented in \cite{BA1}.

In order to provide a  theoretical framework for our findingsin the previous section, we will
consider an analytical model for a dipole cascade. We will assume that
there  is  a  distribution,  $f(\mu,\kappa)d\mu$, that  describes  the
dipoles    with    size    $\mu$    at   the    scale    $\kappa\equiv
\ln(k_{\perp}^2/\Lambda^2)$   (where  $k_{\perp}$   is   the  ordering
variable  and $\mu  \equiv  \ln(M_d^2/k_{\perp}^2)$, in  terms of  the
dipole masses $M_d$). We will consider  the change in $f$ when we make
a small change in the scale $\kappa$.  We note that $\mu + \kappa = ln
\frac{M_d^2}{\Lambda^2}$ is independent  of $\kappa$. If dipoles never
decayed, the  distribution $f(\mu,\kappa)$ would not change  if $\mu +
\kappa$ is kept constant as  $\kappa$ varies.  Therefore any change in
the distribution  along the lines $\mu+\kappa =  constant$ must purely
come from decays.

There are two contributions to this.  A dipole of size $\mu$ may decay
into two smaller dipoles. A dipole of size bigger than $\mu$ may decay
such  that one of  the daughter  dipoles is  of size  $\mu$. Gathering
these terms  together we  get a partial  integro-differential equation,
cf.    \cite{BA1,BA},  based   upon  Eq.    \ref{1gluon}  with
$\alpha_{\mathrm{eff}}=\frac{\alpha_0}{\kappa}$ for gluonic dipoles ($\alpha_0=\frac{6}{11-2N_f/N_c}$
 ):
\begin{eqnarray}
\label{dipole2}
D  f  \equiv   \frac{\partial  f}{\partial  \kappa}  -  \frac{\partial
  f}{\partial  \mu} =  \frac{\alpha_0}{\kappa}\left[\mu f(\mu,\kappa)-
  2\int_{\mu} d\mu^{\prime} f(\mu^{\prime},\kappa)\right]
\end{eqnarray}
The factor of 2 in the contribution from the larger dipoles stems from
the fact  that we are  only considering ``central'' or  purely gluonic
dipoles here.   When a gluonic  dipole decays, there are  2 equivalent
ways to obtain a gluonic dipole of size $\mu$. \\

This  integro-differential  equation  can  be made  into  two  coupled
 differential  equations in  terms of  the first  two moments  of $f$,
 $N_1$ and $N_2$
\begin{eqnarray}
\label{moments12}
N_j(\mu,\kappa)&=&  \int_{\mu}  d\mu_1  (\mu_1)^{j-1}  f(\mu_1,\kappa)
\nonumber  \\  D  N_1  &=& -\frac{\alpha_0}{\kappa}(N_2-  2  \mu  N_1)
\nonumber\\ D N_2 &=&- N_1 +\frac{\alpha_0}{\kappa} \mu^2 N_1
\end{eqnarray}
For the normalisation we note that $N_1(0,\kappa)$ is the total number
 of dipoles available at the scale $\kappa$, to be called $\bar{n}$ in
 this section.  $N_2(0,\kappa)$ is their combined length, to be called
 $\bar{\ell}$.   The equations  for these  quantities can  be obtained
 directly from Eqs. (\ref{moments12}) in the $LLA$-scheme:
\begin{eqnarray}
\label{moments123}
\frac{d  \bar{n}}{d   \kappa}&=&  -\frac{\alpha_0}{\kappa}  \bar{\ell}
\nonumber \\ \frac{d \bar{\ell}}{d\kappa}&=&-\bar{n}
\end{eqnarray}
These equations are solvable in  terms of combinations of the modified
Bessel functions (cf. the next  subsection) as soon as we have defined
a ``starting value'', $\kappa_{max} \equiv L_0$, for the cascade.

In order to solve the  coupled equations in Eq.  (\ref{moments12}), we
 introduce  the  combination  $g(\mu,  \kappa)=  N_2-\mu  N_1$,  which
 fulfils the equation:
\begin{eqnarray}
\label{diffD2}
Dg &=&\frac{\alpha_0}{\kappa}(\mu g)
\end{eqnarray}
with the boundary value
\begin{eqnarray}
g(0,x) &=& \bar{\ell}(x)
\end{eqnarray}
It         is         related         to         $N_1$         through
$N_1(\mu,\kappa)=-\frac{\partial}{\partial \mu}g(\mu,\kappa) $.

Including the boundary condition above, we obtain:
\begin{eqnarray}
\label{solution1}
N_1(\mu,\kappa)&=&            -\frac{\partial}{\partial           \mu}
\left\{\bar{\ell}(\mu+\kappa)  \exp\left[-  \int_{\kappa}^{\mu+\kappa}
dy \frac{\alpha_0}{y}(\mu + \kappa -y)\right] \right\}
\end{eqnarray}

This  is  a derivative  of  two  contributing  factors. The  first  is
$\bar{\ell}$ evaluated  at $\mu +\kappa  \equiv \ln(M_d^2/\Lambda^2)$,
i.e. at  the largest $\kappa$ value  where the dipole  could have been
produced. It is multiplied by  the probability that the dipole has not
decayed  until $\kappa$.   In field  theoretical language  this second
contribution, the Sudakov  form factor, corresponds to the  sum of all
the virtual  corrections during the ``lifetime'' of  the dipole.  This
Sudakov form factor can be easily calculated
\begin{eqnarray}
\label{Sudakov}
S(\mu,\kappa)&=&        \exp\left[-\int_{\kappa}^{\mu+\kappa}       dy
\frac{\alpha_0}{y}(\mu          +          \kappa          -y)\right]=
\frac{\exp(\alpha_0\mu)}{(1+\frac{\mu}{\kappa})^{\alpha_0(\mu+\kappa)}}
\end{eqnarray}

It fulfils
\begin{eqnarray}
\label{Sud2}
S(0,\kappa)      &=&      1      \\      \nonumber      \frac{\partial
S}{\partial\mu}(\mu,\kappa)&=&-\alpha_0 \ln(1+\frac{\mu}{\kappa})S
\end{eqnarray}
The derivative vanishes at $\mu=0$.

The  distribution $f$  is  obtained by  a  partial differentiation  of
$N_1$, cf. Eq. (\ref{moments12})
\begin{eqnarray}
f(\mu,\kappa)=-\frac{\partial N_1}{\partial \mu}
\end{eqnarray}
so   that  using   Eqs.   (\ref{moments123}),   (\ref{solution1})  and
(\ref{Sud2}) we obtain:
\begin{eqnarray}
\label{solf}
f(\mu,\kappa)&=&[\bar{\ell}(L_0,\mu+\kappa)
\alpha_0\ln(1+\mu/\kappa)^2+   \nonumber   \\   &   &2\bar{n}(L_0,\mu+
\kappa)\alpha_0\ln(1+\mu/\kappa)] S(\mu,\kappa)
\end{eqnarray}

The  result  can  be understood  in  a  simple  and useful  way.   The
logarithmic factors in the parenthesis are equal to
\begin{eqnarray}
\label{breakuppoints}
\alpha_0\ln(1+\mu/\kappa)  = \int_{\kappa}^{\mu+\kappa} \frac{\alpha_0
dx}{x}
\end{eqnarray}
i.e. the  probability that  there will be  a dipole  breakup somewhere
between $\mu+ \kappa$, the largest virtuality where a dipole with size
$\mu$  at  $\kappa$  can  be  produced,  and  $\kappa$,  where  it  is
found. There are  two such factors multiplying all  interior points in
the dipoles at $\mu+\kappa$ , counted by $\bar{\ell}(\mu+\kappa)$, and
there  is  one factor  multiplying  twice  the  number of  dipoles  at
$\mu+\kappa$ (because if there is already a dipole one can keep either
its ``left'' side or its ``right'' side and produce the ``other'' side
of the dipole  with size $\mu$ by a breakup  in between). Finally, the
Sudakov form factor  describes the probability that there  is no decay
affecting the existence of the dipole at $\kappa$.

In  the limit  when  $L_0 \gg  (\mu  + \kappa)$,  and $\mu+\kappa$  is
significantly positive,  the two functions  $\bar{n}$ and $\bar{\ell}$
factorise and we may write in the $LLA$
\begin {eqnarray}
\bar{n}(L_0,      x)      &\simeq&      V(L_0)      (x)^{-\frac{1}{4}}
\exp(-2\sqrt{\alpha_0  x})  \\  \nonumber  \bar{\ell}(L_0,x)  &\simeq&
\sqrt{x/\alpha_0} \bar{n}(L_0,x)
\end{eqnarray}
\begin{figure}[t]
\begin{center}
\rotatebox{270}{\epsfig{figure=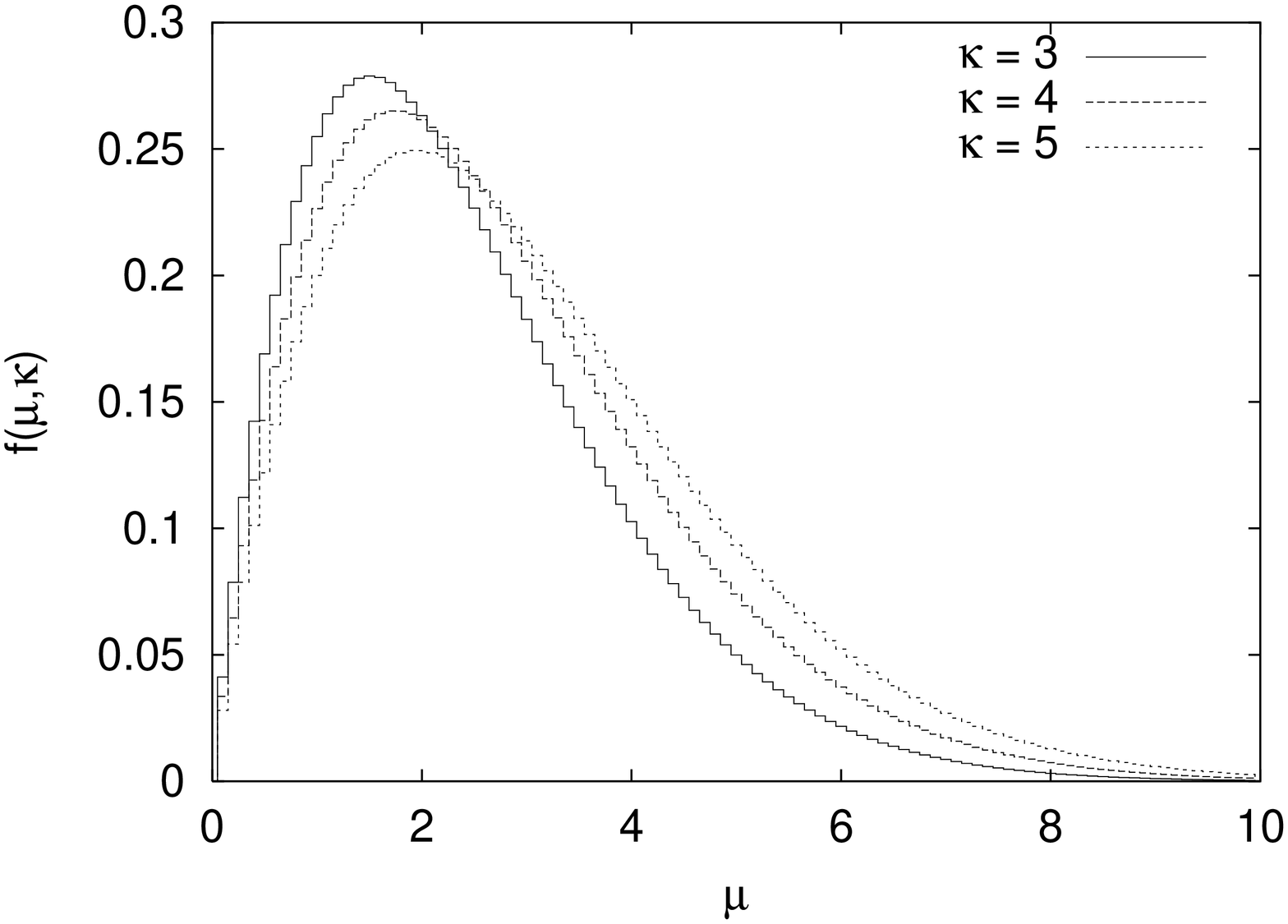,
width=0.5\textwidth }}
\caption{The $f_{\kappa}(\mu)$ function, obtained from Eq. \ref{solf},
normalised to unity.}
\label{figuref}
\end{center}
\end{figure}
In  Fig.   \ref{figuref},  we  show  the  function  $f_{\kappa}(\mu)$,
obtained from $f$ in Eq.  (\ref{solf}) by normalising it to unity, for
a set of values of the  parameter $\kappa$. There is a weak dependence
upon the ordering variable $\kappa$  and for large $\mu$-values it has
an essentially  exponential falloff with  a slope very similar  to the
ones  we  find  in   the  distributions  $P_{\ell}(\Delta  \ell)$  and
$P_{\lambda}(\Delta \lambda)$ in Section \ref{results}.

For  small values of  $\mu$ there  are quantitative  differences.  The
reason is that  the $LLA$ is not sufficient  to describe the behaviour
of small  dipoles. In that  case it is  necessary to include  both the
contributions from the $(\sum_p)$ term  in the dipole cross section in
Eq.   (\ref{1gluon}),  and  also  to  take  into  account  the  recoil
contributions which are particularly  noticeable for small dipoles. We
will consider this question in the next subsection.

\subsection{The $MLLA$ and Earlier Investigations}
\label{MLLA}

In  the  so-called  Modified  Leading  Log  Approximation,  subleading
effects from the polaristaion sum are included \cite{DKMT,GG1}.
In  general, the assumptions  have been  that the  cms energy  $\sqrt{s}$ (or
 rather its  logarithm $L\equiv  \ln(s/\Lambda^2)$) is very  large and
 that  the  main  contributions  from  the  cascades  will  come  from
 well-ordered emissions,  i.e.  that  $k_{\perp j} \gg  k_{\perp j+1}$
 (at least for adjacent emissions).

Using  this approximation,  it  is possible  to  write out  analytical
 equations  for the  expected changes  in  the average  $\ell$ (to  be
 called   $\bar{\ell}(\kappa)$)  and   the   average  $n$   (similarly
 $\bar{n}(\kappa)$)  for  a   given  ordering  variable  $\kappa\equiv
 \ln(k_{\perp}^2/\Lambda^2)$ \cite{DKMT,BA,GG1}:
\begin{eqnarray}
\label{difflan}
\frac{d\bar{\ell}}{d\kappa}       &=&       -\bar{n}       \nonumber\\
\frac{d\bar{n}}{d\kappa}&=& - \alpha_{\mathrm{eff}}(\bar{\ell} - \bar{n}\delta)
\end{eqnarray}
Eqs. (\ref{difflan})  are simplified versions of  the Modified Leading
Log Approximation  (MLLA) formulae. A more  elaborate treatment should
include  the  difference  between  the quark  and  antiquark  endpoint
dipoles with  a different effective coupling $\alpha$  and a different
correction $\delta$,  cf. \cite{GG1}, but these  are small corrections
if the starting value $L=\ln(s/\Lambda^2)$ is large.

The interpretation  is that in a step  $d\kappa$, $\bar{\ell}$ changes
by  $\bar{n}$ contributions  from the  scale change  according  to Eq.
(\ref{ngluonphase}),    just   like   the    second   line    in   Eq.
(\ref{moments123}).  Further, the change  in $\bar{n}$ in that step is
given by  the coupling  times the available  phase space.   This phase
space is given by $\bar{\ell}$  to first approximation; but there is a
correction   from   each   dipole    from   the   term   $\sum_p$   in
Eq. (\ref{1gluon}).   The polarisation sum $\sum_p \simeq  2$ near the
centre of a dipole but decreases in the neighbourhood of the emitters.
It has  been shown  \cite{GG1} that the  suppression of emissions
because of  this can be approximated  by (and this  is essentially the
MLLA)  a constant  decrease in  the dipole  size $\delta  =  11/6$ for
gluonic dipoles (this number only  depends upon the triple
gauge boson vertex for 3 colours). In Eqs. (\ref{difflan}) it is
also assumed that the  gluon splitting process $g\rightarrow q\bar{q}$
is    neglected    and    therefore   $\alpha_{\mathrm{eff}}    \equiv
\alpha_0/\kappa$ with  $\alpha_0 =6/11$ for  all dipoles and  not only
the purely gluonic ones.

The decay  of one  dipole also  affects the two  adjacent ones  due to
recoils.   A   method  to  estimate  this  effect   was  suggested  in
\cite{PE}. Therefore an extra ``loss-term'' is added to the first line
in Eq. (\ref{difflan}):
\begin{eqnarray}
\label{patrik}
\frac{d\bar{\ell}}{d\kappa}= -\bar{n} - C_r \frac{d\bar{n}}{d\kappa}
\end{eqnarray}
with a constant recoil correction estimated to be $C_r\simeq 2$.

These equations can be solved  using the following combinations of the
modified Bessel functions:
\begin{eqnarray}
\label{bessel1}
\mathcal{     I}_1(x)     &=&    \sqrt{2}x^{\gamma/2}     I_{\gamma}(2
\sqrt{\alpha_0x}) \nonumber \\ \mathcal{ I}_2(x) &=& \sqrt{2 \alpha_0}
x^{(\gamma-1)/2} I_{\gamma-1}(2\sqrt{\alpha_0x})
\end{eqnarray}
and similarly  for $\mathcal{K}_1$  and $\mathcal{K}_2$ in  terms of
the  exponentially falling  Bessel function  $K$. Note  that  they are
normalised so that
\begin{eqnarray}
\label{bessel2}
\mathcal{ I}_1(x)\mathcal{ K}_2(x)+ \mathcal{ I}_2(x)\mathcal{ K}_1(x)
=x^{\gamma-1}
\end{eqnarray}
and that there are simple differential relations between them:
\begin{eqnarray}
\label{bessel3}
\frac{d \mathcal{ I}_1}{dx}  &=& \mathcal{ I}_2(x) \nonumber\\ \frac{d
\mathcal{           I}_2}{dx}&=&           \frac{\alpha_0}{x}\mathcal{
I}_1(x)+\frac{(\gamma-1)}{x}\mathcal{ I}_2(x)
\end{eqnarray}
Similar relations  hold for the exponentially  falling modified Bessel
function pair $(\mathcal{ K}_1,-\mathcal{ K}_2)$.

If we start the cascade  at a large value of $L\equiv \ln(s/\Lambda^2)
\gg  \kappa$ we obtain,  with the  boundary values  $\bar{\ell}=0$ and
$\bar{n}=1$ at $L$, the results:
\begin{eqnarray}
\label{solvelan}
a&=&  L^{1-\gamma}\left[C_r\mathcal{  K}_2(L)-\mathcal{ K}_1(L)\right]
\\  \nonumber  b&=&  L^{1-\gamma}\left[C_r\mathcal{  I}_2(L)+\mathcal{
I}_1(L)\right]      \\       \nonumber      \bar{n}&=&      b\mathcal{
K}_2(\kappa)-a\mathcal{   I}_2(\kappa)  \\   \nonumber  \bar{\ell}+C_r
\bar{n}&=&a\mathcal{ I}_1(\kappa)+b\mathcal{ K}_1(\kappa)
\end{eqnarray}
using the notations $\gamma =1 +\alpha_0(\delta+C_r)$.

It   is  difficult   to   obtain  an   anlytically  solvable   partial
integro-differential      equation      corresponding      to      Eq.
(\ref{dipole2}). The method of  moments runs into difficulties because
the  variation  in  the  $\sum_p$  term  in  the  cross  section  (Eq.
(\ref{1gluon})) affects the integral contribution, i.e.  the gain from
the decays of larger dipoles, in such a way that the equations for the
first  few  moments  are  no  longer independent  of  higher  moments.
Therefore,  we  will   directly  go  over  to  the   solution  in  Eq.
(\ref{solf}) instead,  and modify the  ingoing terms according  to the
MLLA, keeping in mind the interpretation given there.

We  assume that  each  dipole of  size  $\mu$ can  decay  only in  the
interior  excluding a  region  $\delta/2$ on  each  extreme. Then  the
effective  phase  space  at  $(\mu+\kappa)$ in  Eq.   (\ref{solf})  is
changed  from $\ell(\mu+\kappa)$ into  $[\ell(\mu+\kappa)-\delta \cdot
n(\mu+\kappa)]$.   A dipole  of size  $\mu$ is  produced by  a breakup
first at  a point $\kappa=\kappa_1$ and  then at $\kappa_2<  \kappa_1$. The region
where  the ``no  emission probability''  is  to be  calculated in  the
Sudakov  form factor  is  changed in  the  MLLA, first  by  a loss  of
$\delta/2$  at $\kappa_1$  and then  by  a loss  of a  further $\delta/2$  at
$\kappa_2$.
\begin{eqnarray}
\label{mllasudakov}
S_{\kappa_1,\kappa_2}&=&               \exp\left(-\int_{\kappa}^{\kappa_2}
\frac{\alpha_0                       dz(\mu+\kappa-\delta-z)}{z}\right)
\exp\left(-\int_{\kappa_2}^{\kappa_1}
\frac{\alpha_0dz(\mu+\kappa-\delta/2-z)}{z}\right) \times \nonumber \\
&                                     &\exp\left(-\int_{\kappa_1}^{\mu+\kappa}
\frac{\alpha_0dz(\mu+\kappa-z)}{z}\right)        =\left(       \frac{\kappa_1
\kappa_2}{\kappa^2}\right)^{\frac{\alpha_0     \delta}{2}}    S(\mu,
\kappa)
\end{eqnarray}
where  $S(\mu,\kappa)$ is  the Sudakov  form factor  without  the term
 $\delta$ as given  in Eq. (\ref{Sudakov}).  Since this  is a function
 of  $\kappa_1$ and  $\kappa_2$ now,  we  obtain, for  the contribution  to
 $f(\mu,  \kappa)$ from  situations where  both gluons  involved  in a
 dipole are emitted below $\mu+ \kappa$, the result:
\begin{eqnarray}
\label{solf2}
f_{1}&\equiv&         2(\bar{\ell}-\delta         \bar{n})(\mu+\kappa)
\int_{\kappa}^{\mu+\kappa}   \frac{\alpha_0  d\kappa_1}{\kappa_1}  \int_{\kappa}^{\kappa_1}
\frac{\alpha_0d\kappa_2}{\kappa_2}S_{\kappa_1,\kappa_2}   \nonumber  \\
&=&(\bar{\ell}-\delta
\bar{n})(\mu+\kappa)\frac{4\left((1+\mu/\kappa)^{\frac{\alpha_0
\delta}{2} }- 1\right)^2}{\delta^2} S(\mu,\kappa)
\end{eqnarray}
The fact that the first break may  be either on the ``left'' or on the
``right'' side contributes a factor of $2$ in the above equation .
\begin{figure}[t]
\begin{center}
\mbox{
\hspace{-0.5cm}
\rotatebox{270}{\epsfig{file=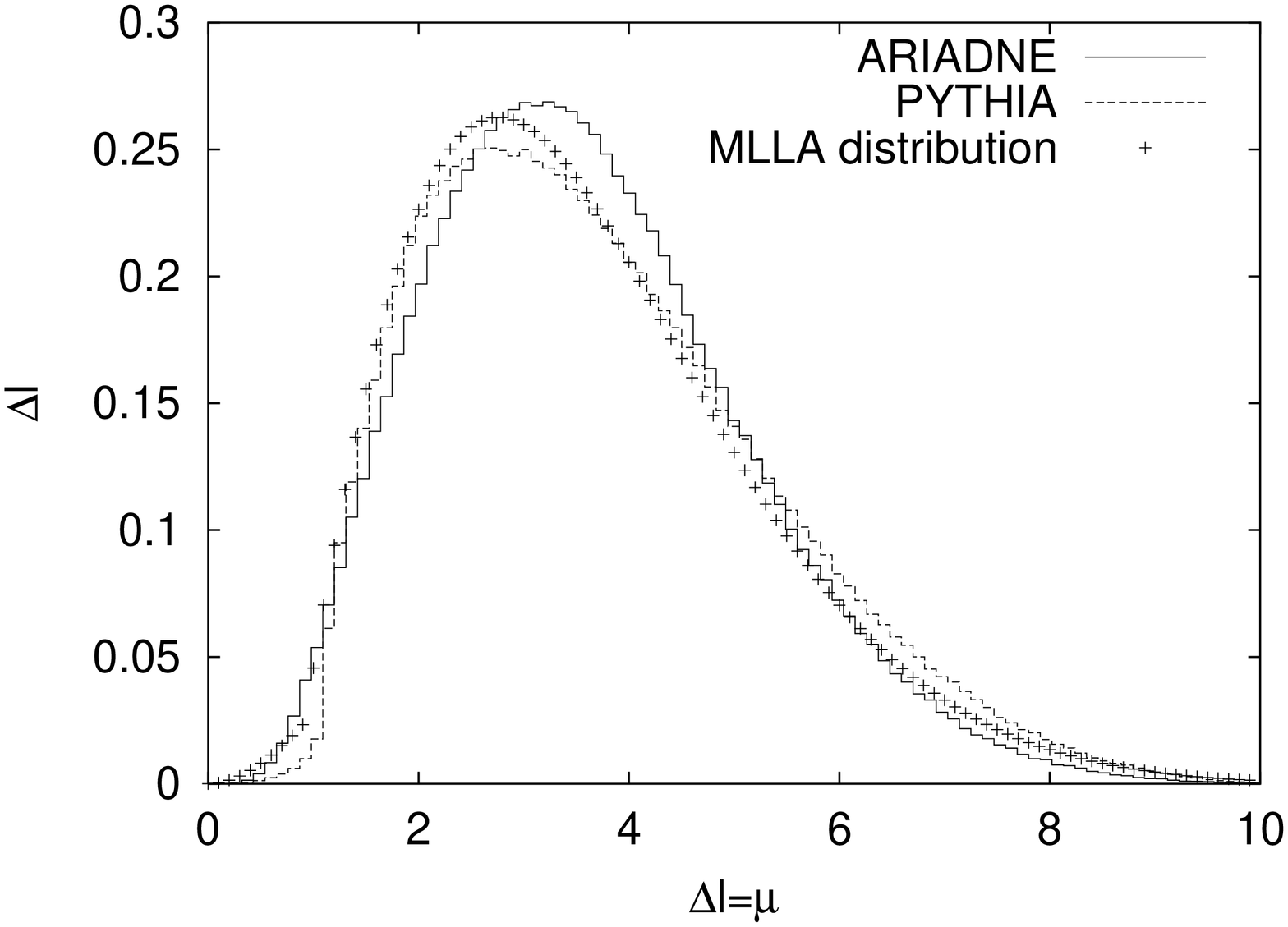,width=6.0cm}}
\hspace{-0.5cm}
\rotatebox{270}{\epsfig{file=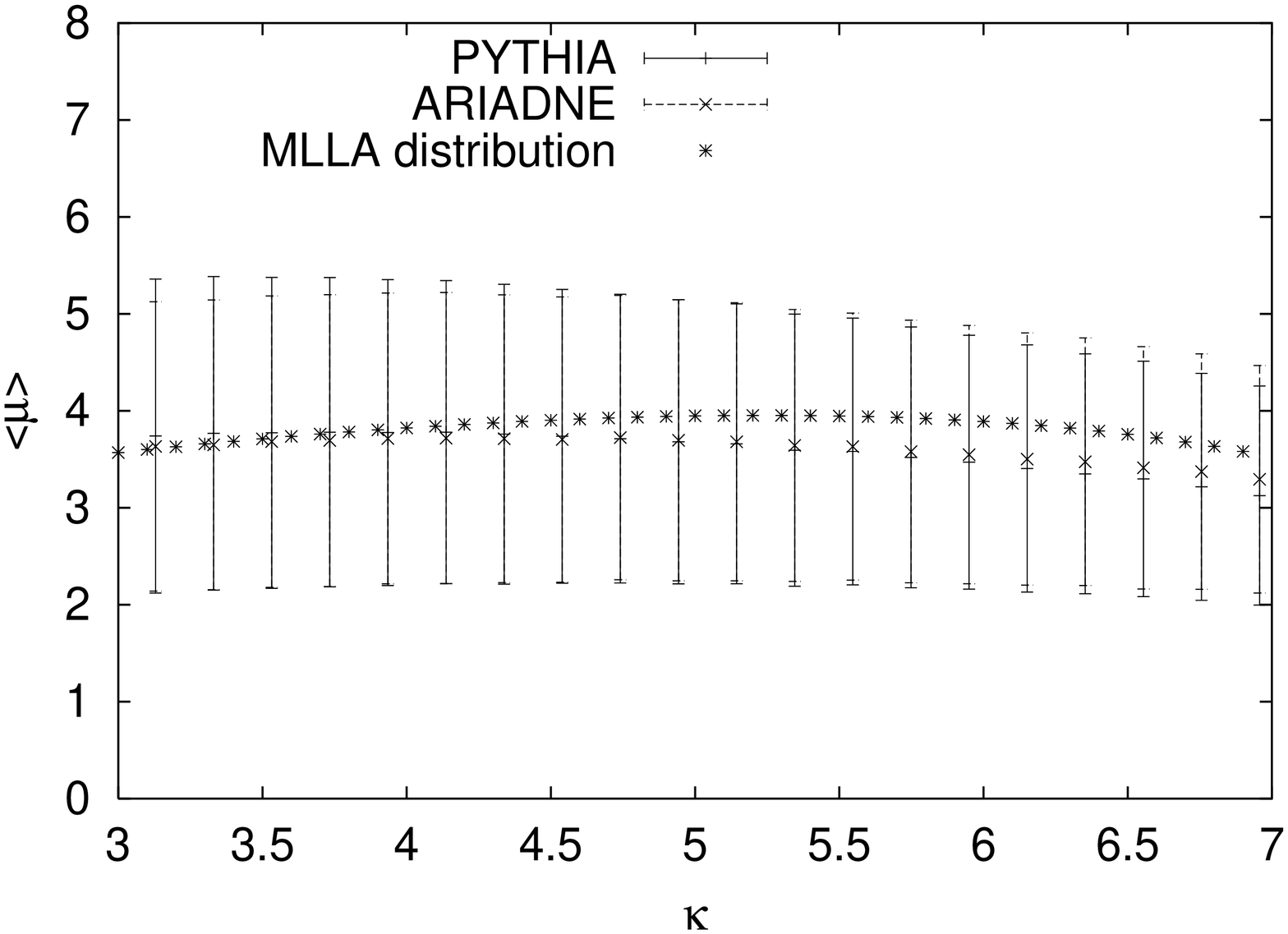,width=6.0cm}}      }
\captive{The   figure    to   the   left    shows   the   distribution
$f_\delta(\mu,\kappa)$,   obtained   from   the   MLLA   approach   at
$k_\perp=$$1GeV$,   compared    with   the   $P_{\ell}(\Delta   \ell)$
distribution from  {\Ar} and {\Py} (cf.   Fig.  \ref{jetsetDela}). The
figure to the right shows the  average value of $\mu$ as a function of
$\kappa$, from  the same distribution  $f_\delta(\mu,\kappa)$.  In the
figure to the right, we also  compare with the average value of $\ell$
(cf.  Eq.   (\ref{sizektlambda})) as  a function of  $\kappa$ obtained
from {\Ar}  and {\Py}, and we  also include the  standard deviation in
the  result  from {\Py}  and  {\Ar}.  This  result is  obtained  using
$\alpha_0=0.7$ and  $C_r=2$ in the  MLLA distribution as  explained in
the text.
\label{diffgraph}
}
\end{center}
\end{figure} 

In the same way, for the  second contribution related to the number of
dipoles  at  $\mu+\kappa$, ie  the  situations  where  only one  gluon
involved  in a  dipole is  produced below  the scale  $\mu+\kappa$, we
obtain,
\begin{eqnarray}
\label{solf3}
f_2         =         \theta(\mu-\delta/2)         \bar{n}(\mu+\kappa)
\frac{4\left((1+(\mu-\delta/2)/\kappa)^{\frac{\alpha_0       \delta}{2}
}-1\right)} {\delta}S(\mu-\delta/2,\kappa)
\end{eqnarray}
In this  case there can  never be a  breakup until the dipole  size has  at
least reached $\delta/2$, because  one of the end points is already
fixed at the virtuality $\mu+\kappa$.

We note  for consistency that  the result $f_{\delta}  \equiv f_1+f_2$
has the  limit $f$  in Eq.  (\ref{solf})  when $\delta$  vanishes.  In
Fig. \ref{diffgraph}, we show the result of this $MLLA$ version of the
dipole size distribution. We have used a fixed value $\delta=11/6$ but
we have allowed the parameters  $\alpha_0$ and $C_r$ vary.  The reason
for allowing $\alpha_0$ to vary is that it is feasible to shut off the
gluon splitting process in {\Ar}, but the running coupling will always
receive contributions from the different number of ``active'' flavours
at different  virtualities. From the  figure to the left,  we conclude
that it is  in between the distributions from {\Ar}  and {\Py} for the
central dipoles when we use $\alpha_0 = 0.70$ (i.e.  for three to four
active flavours),  and $C_r=2$ (according to  \cite{PE}).  This result
is rather insensitive  to $C_r$ and it can therefore not  be used as a
test for the claims in \cite{PE}.  We have also checked the dependence
upon the ordering variable.  In Fig.  \ref{diffgraph} to the right, we
show  that $f_{\delta}$  also exhibits  a slow  change of  its average
value as a function of the (relevant) ordering variable like the {\Ar}
and {\Py} $P_{\ell}$ distributions.

\section{Concluding Remarks}
\label{conclu}
The  number of  gluons  emitted {\it  above}  a certain  value of  the
ordering  variable  shows  an  interesting  linear  correlation  (with
gaussian fluctuations)  with the phase space available  at that scale.
We have  called the  phase space variable  the $\ell$-measure  in this
paper  to  distinguish it  from  the  $\lambda$-measure,  which is  an
infrared stable  generalisation of  the $\ell$-measure, useful  in the
context of hadronisation.

Both these quantities  are functions of not only  the resolution scale
($k_{\perp  cut}$   for  a   dipole  cascade,  $\Gamma_0$   in  string
fragmentation), but  also of the  precise geometry of the  state under
consideration.  Therefore they change  with every new gluon emitted in
the system.

The  resolution parameter  relevant for  hadronisation is  a parameter
fixed by the  break-up properties of the string  and the mass spectrum
of the hadrons produced.  The $\lambda$-measure is constructed in such
a way that addition of  gluons at transverse momenta much smaller than
the scale  $\Gamma_0$ does  not change  its value. In  that way  it is
infrared stable.

However, since the resolution  parameter for the $\ell$-measure is the
transverse momentum of the gluon to be emitted, the transverse momenta
of  all gluons already  emitted must  be higher  in an  ordered dipole
cascade.  Therefore, the only  kind of infrared stability relevant for
this measure would be the stability, when the ordering variable, which
is also  the resolution parameter,  is small. Technically this  is not
the case. However, since in  this approach we will never compute phase
space  available  at  a  value  of the  ordering  variable  below  the
$\Lambda_{QCD}$,  there is  an  effective infrared  stability for  the
phase space, if all local  transverse momenta $k_{tj}$ are required to
be above $k_{\perp}$.  It is easily shown that the  mass of any dipole
with  a gluon  at  one end  is  greater than  the transverse  momentum
$k_{t}$  of  that   gluon,  which  in  turn,  will   be  greater  than
$\Lambda_{QCD}$. Besides,  we must  remember that it  is kinematically
not possible for dipoles of mass smaller than $2\Lambda_{QCD}$ to emit
gluons above $\Lambda_{QCD}$.

The two  measures discussed here  are quite strongly related.   If one
keeps only the  last (often the dominant) term in  the argument of the
logarithm  in  the  $\lambda$-measure,  one gets  an  expression  very
similar  to the  $\ell$-measure.
    
The linear  correlation between the  $\ell$-measure and the  number of
gluons is  a direct  consequence of the  following facts that  we have
demonstrated.   The $\ell$-measure  is the  sum of  a series  of terms
$\Delta \ell_j$, one  for each dipole. A given  $\Delta \ell_j$ is the
contribution from one particular  dipole to the $\ell$-measure, and is
in some  sense the ``size''  of that dipole.   There are two  kinds of
dipoles. The  first kind  involves either the  quark or  the antiquark
energy--momentum.    The   second   kind   involves  only   the   gluon
energy--momenta.   These   two   kinds   of  dipoles   have   different
distributions in sizes.

It turns out that the size  of the dipoles do not show any significant
correlations. Therefore the $\ell$-measure  of a state consisting of a
certain number of dipoles, which  is just the sum of the corresponding
$\Delta \ell_j$'s, is distributed  like the sum of several independent
random variables, i.e.  like a gaussian with a mean that is the sum of
the means of the  individual independent distributions.  This shows up
as  a  linear  correlation  between  the number  of  partons  and  the
$\ell$-measure.

More  interesting than  the linear  relation is  the stability  of the
slope of these  lines, which represents the number  of gluons per unit
phase space, with  respect to changes in global  event parameters like
cms  energy  or   thrust  and  even  with  changes   in  the  ordering
variable. It seems that parton cascades tend to form structures at one
(relative) size when sizes are measured with the $\ell$-measure with a
resolution parameter proportional to  the ordering variable.  As we go
down in  the ordering variable,  the running coupling  should increase
the  number of  emissions per  unit phase  space.  However,  the phase
space variable  itself scales appropriately to  effectively absorb any
increase of emissions per unit phase space.

We have  also made  another observation in  this paper. Just  like the
$\ell$-measure, the  $\lambda$-measure can be  thought of as a  sum of
terms $\Delta \lambda_j$,  one for each vector along  the directrix in
colour order. Out of the  two vectors involved in a particular $\Delta
\lambda_j$,  only one  is  a  vector along  the  directrix. The  other
vector, in a sense a cumulative variable, is a weighted average of the
vectors $k_i$'s along the directrix with $i<j$.

It is  interesting that  all the properties  of the  $\ell$-measure we
have discussed here are also reflected in the $\lambda$-measure. There
is a similar linear correlation  to the number of gluons, similar lack
of correlation among adjacent terms and similar stability with respect
to   change    of   global   event   variables    and   the   ordering
variable. Partitioning the $\lambda$-measure  into a series of $\Delta
\lambda_j$'s corresponds  to partitioning the  surface spanned between
the  directrix   and  the  $\mathcal{  X}$-curve  into   a  series  of
``plaquettes'' or flat regions  along the directrix. Each plaquette is
spanned  between  a parton  energy  momentum  vector  $k_j$, a  vector
$q_{Tj-1}$  (weighted  average of  $k_i$'s  along  the directrix  with
$i<j$),  another  vector $q_{Tj}$  and  a  hyperbolic  section of  the
$\mathcal{ X}$-curve.  The size of the plaquette $\Delta \lambda_j$ is
the  length  of  the  section  of the  $\mathcal{  X}$-curve,  and  is
determined by  $k_j$ and $q_{Tj-1}$.  Since these  plaquettes share so
many  properties  with  dipoles,   we  will  call  them  ``Generalised
Dipoles''.

We  note   in  passing  that  the  possibility   of  partitioning  the
contributions to the $\lambda$-measure as contributions from connected
flat regions has an interesting consequence for the fragmentation of a
string  according to  the Lund  model area  law. Such  a fragmentation
scheme was  presented in \cite{BASMFS},  which was closely  related to
the   $\lambda$-measure  and   the  $\mathcal{   X}$-curve.   Particle
production  in this  scheme  can  be thought  of  as partitioning  the
plaquettes  mentioned  here  into  smaller plaquettes,  one  for  each
hadron. Since  the hadron energy--momenta  will mostly come  from these
flat regions in  the string, particles stemming from  one GD will have
energy--momenta aligned in a plane in Minkowski space, up to transverse
momentum  fluctuations.  We have  found such  chains of  particles and
examined  their   properties.   These  chains,  which   we  will  call
``coherence  chains'' are  important for  the study  of  Bose Einstein
correlations in the  Lund Model for multigluon systems,  which we will
elaborate on in a forthcoming publication.

We have also  presented here an analytical model  for the observations
we have  made about dipoles, in  the LLA and MLLA  schemes.  There are
limitations  to  this  approach,   since  complications  such  as  the
polarisation  sum  make  it  very  difficult  to  obtain  differential
equations which can be solved analytically. Nevertheless, our analysis
based  on  simple  gain-loss  considerations  broadly  reproduces  the
qualitative features of the observations.

\section{Acknowledgements}
We would like to thank G. Gustafson, T.  Sj\"ostrand and L. L\"onnblad
 for helpful discussions.  This work  was completed by S.  Mohanty and
 F.   S\"oderberg, about  two months  after the  untimely death  of our
 collaborator Bo Andersson.


\end{document}